\documentclass{aa}
\usepackage{graphicx}
\usepackage{txfonts}
\usepackage{longtable} 
\usepackage{natbib} 
\usepackage[latin1]{inputenc}

\begin{document}

\title{Age and metallicity of star clusters in the Small Magellanic Cloud
from integrated spectroscopy}
\author{
B. Dias\inst{1}
\and P. Coelho\inst{2}
\and B. Barbuy\inst{1}
\and L. Kerber\inst{1,3}
\and T. Idiart\inst{1}
\fnmsep
}

\institute{Universidade de S\~ao Paulo, Dept. de Astronomia, Rua do Mat\~ao 
1226, S\~ao Paulo 05508-090, Brazil\\
email: bdias, barbuy, idiart, kerber@astro.iag.usp.br
\and
N\'ucleo de Astrof\'{\i}sica Te\'orica - Universidade Cruzeiro do Sul,
R. Galv\~ao Bueno, 868, sala 105, Liberdade, 01506-000,
S\~ao Paulo, Brazil \\
email: paula.coelho@cruzeirodosul.edu.br
\and
Universidade Estadual de Santa Cruz, Rodovia Ilh\'eus-Itabuna km16, 45662-000 Ilh\'eus, Bahia, Brazil
}


\abstract
{Analysis of ages and metallicities of star clusters in the Magellanic
  Clouds provide information for studies on the chemical evolution of
  the Clouds and other dwarf irregular galaxies.} 
{The aim is to derive ages and metallicities from integrated spectra of
14 star clusters in the Small Magellanic Cloud,
including a few intermediate/old age star clusters.
}
{Making use of a full-spectrum fitting technique,
we compared the integrated spectra of the sample clusters to 
three different sets of single stellar population 
models, using two fitting codes available in  the literature. }
{We derive the ages and metallicities of 9 intermediate/old age clusters, some
of them previously unstudied, and 5 young clusters. }
{We point out the interest of the newly identified as intermediate/old age 
clusters HW1, NGC 152, 
Lindsay 3, Lindsay 11, and Lindsay 113. We also confirm the old ages of NGC 361, NGC 419,
 Kron 3, and of the very
well-known oldest SMC cluster, NGC 121.}
\titlerunning{Age and Metallicity of Star Clusters in the  SMC}
\maketitle


\section{Introduction}

The Small Magellanic Cloud (SMC) is classified as a
dwarf irregular galaxy (dI), with an absolute magnitude of M$_{\rm
V}$ $\approx$ -16.2 (\citealp{binney98}), among a variety of other
types of dwarf galaxies in the Local Group (see \citealp{tolstoy09}). 
The star formation history and chemical evolution of the SMC can only be 
understood by deriving ages and
metallicities of its stellar populations. 

Globular clusters in the Milky Way are all old, and  the Large 
Magellanic Cloud (LMC) shows a well-known age gap, with no clusters between
the ages of 4 and 10 Gyr (except for ESO 121-SC03), as first revealed
by \cite{jensen88} and reconfirmed since then (e.g. \citealp{balbinot}
and references therein). Therefore the
SMC appears as a unique nearby dwarf galaxy that has star
clusters of all ages and a wide range of metallicities; yet,they
 are still poorly studied.
For example, Parisi et al. (2009) point out that their spectroscopic
study of CaII triplet metallicities of 16 SMC clusters is the largest
spectroscopic survey of these objects in the SMC.

Ages and metallicities of star clusters in external galaxies beyond the Local Group
 can only be derived from their integrated spectra with current observing
 facilities. In contrast, due to its proximity, the stellar populations of
the Magellanic Clouds (MCs) can be studied through resolved
 colour-magnitude diagrams (CMDs) and spectroscopy 
of bright individual stars, HII regions, or planetary nebulae.

High-resolution abundance analyses of field individual stars 
in the SMC  in supergiants, among
which the largest samples can be found in \cite{hill97} for K-type stars, \cite{luck98} for
Cepheids, and \cite{venn99} for A-type stars, were shown by Tolstoy et al. (2009)
to be compatible with other dIs of the Local Group.
Spectroscopy of HII regions in the SMC were carried out by \cite{garnett95},
where objects with  7.3 $<$ log (O/H)+12 $<$ 8.4 were
studied. Spectroscopic abundances of a large sample of SMC planetary
nebulae were presented in \cite{idiart07}, where objects with
 6.69 $<$ log (O/H)+12 $<$ 8.51 were presented. 

Ages of star clusters in the SMC using CMDs can now be obtained with improved
angular resolutions, reaching several magnitudes below the
turn-off, with higher accuracy (e.g.,  \citealp{glatt08a}, \citealp{glatt08b}
and references therein). \cite{degrijs08} made use of the UBVR survey
of the Clouds by \cite{massey02}, to derive relative ages and masses
of stars clusters in the SMC, where we see that there are few star
clusters with ages of 1 Gyr and older.
 \cite{hill06} presented structural parameters for 204 star clusters,
and \cite{glatt09} give structural parameters for 23 intermediate/old age
star clusters in the SMC.  Chemical abundances from high-resolution 
spectra, on the other hand, are only available for two globular  clusters, 
the young NGC 330 (Hill 1999;
\citealp{gonzalez99} and references therein) and
the old NGC 121 (\citealp{johnson04}), the latter however only as preliminary work. Other high or
mid resolution spectra are essentially only
found for hot stars in young clusters or  from surveys of the CaT triplet lines
(e.g. Parisi et al. 2009; da Costa \& Hatzidimitriou 1998).

Recent work on the star formation history (SFH) of the SMC, based on large samples
of field stars, have shown that its SFH is rather smooth and well-understood,
as described in \cite{dolphin01}, \cite{harris04},
and \cite{carrera08}.
 \cite{dolphin01} studied the SFH
in a field around the old cluster NGC 121. A broadly peaked
SFH is seen, with a high rate between 5 and 8
Gyr ago.
Constant star formation from $\sim$15 to $\sim$2 Gyr ago could be
adopted as well within 2$\sigma$. 
The metallicity increased from the early value of [Fe/H]$\approx$-1.3
for ages above 8 Gyr, to [Fe/H]=-0.7 presently.

\cite{harris04}, based on a UBVI catalogue
of 6 million stars located in 351 SMC regions (\citealp{1997AJ....114.1002Z}),
found similar evidence regarding the SMC's SFH.
 They suggest the following main characteristics:
a) a significant epoch of star formation with
ages older than 8.4 Gyr; b) a long quiescent epoch between 3 and 8.4 Gyr;
c) a continuous star formation started 3 Gyr until the present;
d) in period c), 3 main peaks of star formation should have occurred at 
2-3 Gyr, 400 Myr, and 60 Myr.
De Grijs \& Goodwin (2008) claim to confirm the findings by \cite{gieles07},
according to which there is little cluster disruption in the SMC, and
therefore star clusters should be reliable tracers of star formation history,
together with field stars.

We present the analysis of 14 star clusters, 
several of which still have very 
uncertain age and metallicity determinations. Our sample includes
candidates to be of intermediate/old age.

We use the full-spectrum fitting codes {\sc Starlight}
  \citep{2005MNRAS.358..363C} and \emph{ULySS} \citep{koleva+09}
to compare, on a pixel-by-pixel basis, the integrated spectrum of the
clusters to three sets of simple stellar population (SSP) models in
order to derive their ages and metallicities. 
This technique is an 
improvement over older methods
(e.g., the Lick/IDS system of absorption line indices, \citealt{worthey+94}) and
has been recently validated in \cite{koleva+08,cid_delgado10} and references therein.
If integrated spectra can be proved
to define reliable ages and metallicities, the technique can then be applied
to other faint star clusters in the MCs, 
and in other external galaxies. As a check of our method, we compare our results
to those from CMD analyses or other techniques available for the sample clusters. 
Previously we observed spectra for 14 clusters (6 in the SMC and 8 in the
LMC), and we analysed them based on single stellar population
  models of integrated colours, as well as of H$\beta$ and $<$Fe$>$
  spectral indices (\citealp{freitas98}).

We selected our sample from
  \cite{sagar89} (L3, K3,  L11 and L113), \cite{hodge74} and Hodge (1983)
  (HW 1, NGC~152, NGC~361, NGC~419, and NGC~458), and the
  well-known NGC~121. Young clusters were also included in the sample
  (NGC 222, NGC 256, NGC 269, and NGC 294) for comparison purposes.
Our main aim is to identify intermediate/old star clusters in our sample. 
We assume that `intermediate/old age' populations are older than the Hyades 
(700~Myr, \citealp{friel95}), as commonly adopted in our Galaxy. 
  
In Sect. 2
we describe the observations and data reduction.
In Sect. 3 we present the stellar population analysis.
In Sect. 4 we discuss the results obtained. In Sect. 5
we comment on each cluster. 
Concluding remarks are given in Sect. 6.


\section{Observations}

Observations were carried out at the 1.60m te\-les\-co\-pe of the Na\-tio\-nal 
La\-bo\-ra\-to\-ry for As\-tro\-phy\-sics (LNA/MCT, Brazil) 
and at the 1.52m telescope of 
the European Southern Observatory (ESO, La Silla, Chile).
At the LNA, an SITe CCD camera of 1024x1024 pixels was used, with pixel size
of 24 $\mu$m. A 600 l/mm grating allowed a spectral resolution
of 4.5 {\rm \AA} FWHM. Spectra were centered at Mgb $\lambda$5170,
including the indices H$\beta$ $\lambda$4861, Fe $\lambda$$\lambda$
5270, 5335, Mg$_2$, and NaD $\lambda$5893.
At ESO, a Loral/Lesser CCD camera \#38, of 2688$\times$512 pixels
 with pixel size of
15 $\mu$m and a grating of 600 l/mm were used, allowing a spectral
resolution of 4 {\rm \AA} FWHM. 

We used long  east-west slits in all observations (3 arcmin at LNA 
and 4.1 arcmin at ESO). For each cluster, from 2 to 6 individual measures 
were taken, mainly covering its brightest region. Integration times range 
from 20 to 50 min. each, and we used slits of 2 arcsecs width in both 
observatories.

We used the IRAF package for data reduction, following
the standard procedure for longslit CCD spectra: correction of bias, 
dark and flat-field, extraction, wavelength, and flux calibration.
For flux calibration, we observed at least three spectrophotometric standard 
stars each night. 
Standard stars were observed in good sky conditions, with wider slits to 
ensure the absolute flux calibration. In some nights weather conditions were 
not photometric; however, even in these nights we observed
spectrophotometric standards in order to secure a relative flux
calibration.
Atmospheric extinction was corrected through mean 
coefficients derived for each observatory.
After reduction we combined the spectra to increase global S/N and
filtered in order to minimise high-frequency noise, lowering their
resolutions to about 7 {\rm \AA} FWHM.
The Table \ref{log} presents the log of observations. Reddening values
E(B-V) given in Table \ref{log} were obtained by using the reddening
maps of \cite{schlegel98}.
Signal-to-noise ratios per pixel measured
on the final spectra are also reported in Table \ref{log}.

\begin{table*}
\centering
\caption{Log of observations.}
\label{log}
\begin{tabular}{lllllllllr}
\noalign{\smallskip}
\hline
\noalign{\smallskip}
{\rm cluster} & diam & {$\alpha$(J2000)} & {$\delta$(J2000)} & t$_{exp}$ & n$_{exp}$ & Date  & Site & E(B-V)$^*$ & S/N @ $\lambda\lambda$ \\
\noalign{\smallskip}
                    & (')       & ($^h$ $^m$ $^s$) & ($^o$ ' '')            & (min)     &                &           &        &           &       \\
\noalign{\smallskip}
\hline
\noalign{\smallskip}
HW1	& 0.95 	& 00:18:25 	& -73:23:38 & 30 	& 4 	& 05.08.99  & LNA & 0.037 & 50 @ 6213-6262 \\
K3 (LNA)	        & 3.40 	& 00:24:46 	& -72:47:38 & 30 	& 2 	& 04.08.99  & LNA & 0.037 & 15 @ 6213-6262 \\
K3 (ESO)  	& 	        &  		        &  	            & 40 	& 3 	& 06.08.00  & ESO &           & 114 @ 6082-6132 \\
L3     	& 1.0  	& 00:18:26 	& -74:19:07 & 30 	& 4 	& 04.08.99  & LNA & 0.038 & 16 @ 6213-6262 \\
L11  	& 1.70 	& 00:27:45 	& -72:46:56 & 30 	& 2 	& 31.12.99  & ESO & 0.037 & 18 @ 5820-5880 \\
L113 	& 4.40 	& 01:49:31 	& -73:44:04 & 30 	& 3 	& 30.12.99  & ESO & 0.047 & 33 @ 5454-5490 \\
NGC 121 (LNA)  	& 3.80      & 00:26:43 	& -71:31:57 & 20 	& 3 	& 03.08.99  & LNA & 0.054 & 66 @ 6213-6262 \\
NGC 121 (ESO)  	&     	&  		        &  	            & 40,40& 3 	& 05.08.00  & ESO &           & 253 @ 6082-6132 \\
NGC 152       & 3.00	& 00:32:55     	& -73:07:04 & 30    & 2  & 30.12.99  & ESO & 0.037 & 25 @ 5940-6010     \\
NGC 222 	& 1.20 	& 00:40:42 	& -73:23:00 & 40 	& 2 	& 07.08.00  & ESO & 0.037 & 69 @ 6086-6126 \\
NGC~256 	& 0.90 	& 00:45:54 	& -73:30:26 & 40 	& 2 	& 08.08.00  & ESO & 0.037 & 175 @ 6213-6262 \\
NGC~269 	& 1.20 	& 00:48:21 	& -73:31:49 & 40 	& 2 	& 09.08.00  & ESO & 0.037 & 130 @ 6213-6262 \\
NGC~294 	& 1.70 	& 00:53:04 	& -73:23:17 & 30 	& 2 	& 31.12.99  & ESO & 0.037 & 90 @ 5820-5880 \\
NGC~361 (ESO99) 	& 2.60 	& 01:02:11 	& -71:36:24 & 30 	& 2	& 31.12.99  & ESO & 0.037 & 43 @ 5820-5880 \\
NGC~361 (ESO00)   	&   	        &   	                &                  & 40 	& 3	& 05.08.00  & ESO &           & 131 @ 6213-6262 \\
NGC~419 	& 2.80 	& 01:08:18 	& -72:53:06 & 40 	& 2 	& 08.08.00  & ESO & 0.037 & 149 @ 6213-6262 \\
NGC~458 	& 2.60 	& 01:14:53 	& -71:33:04 & 50 	& 1 	& 08.08.00  & ESO & 0.037 & 173 @ 6213-6262 \\
\noalign{\smallskip}
\hline 
\end{tabular}
\\ \begin{flushleft}$^*$The reddening values E(B-V) are from maps by Schlegel et al. (1998).\end{flushleft}
\end{table*}


\section{Stellar population analysis}
\label{sec_SP}
\subsection{Literature data}
Literature data available for the sample
clusters are reported in Table \ref{dadosliteratura}. 
Values of reddening E(B-V), metallicity and age are listed,
together with the corresponding references.
The metallicity value is either the iron abundance [Fe/H]
or the overall metallicity relative to solar [M/H] = [Z/Z$_{\odot}$] =
[Z], indicated by superscript 1 or 2, respectively. We assumed
Z$_{\odot}$~=~0.017 for the cases where only
the Z value was provided. Table \ref{dadosliteratura} shows that there is very
little information for most clusters. We can see that ages for some of
them vary by several Gyr.  

\begin{figure}
\begin{center}
\includegraphics[width=0.8\columnwidth]{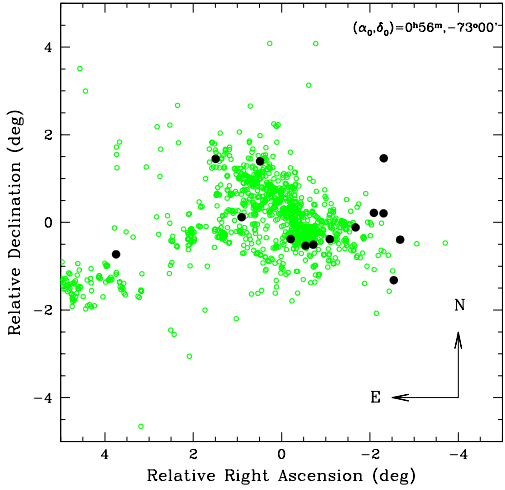}
\caption{Distribution on the sky of the 14 SMC clusters analysed in this
work (large full circles) overplotted on the star clusters listed in
(\citealp{1995ApJS..101...41B})
 (small green circles).}
\label{smcsky}
\end{center}
\end{figure}

The spatial distribution of the 14 sample SMC clusters 
is shown in Fig. \ref{smcsky}, where
the sample clusters are projected over the location of SMC star
clusters as given in \cite{1995ApJS..101...41B}. 
As can be seen in this figure, only one
cluster (Lindsay 113) is located in the direction of the Bridge (east side),
whereas the others are spread over the bulk of the clusters but
 on the west side of the SMC.

In Table \ref{tab_lit} we report the literature values adopted
throughout the text, selected by weighting more reliable methods or
data quality. In this table we also report estimated cluster
  masses. We estimated the masses of the clusters based on the SSP
  evolution models by \cite{schulz02}. In their Fig. 8
  values of V-band mass-to-light ratios (M/L$_V$) are plotted against
  age for three different metallicities ([Z/Z$_{\odot}$]~=~-1.63,
  -0.33, 0.07).  The M/L$_V$ ratios were estimated by considering
  IMF prescriptions  by either \cite{salpeter55} or \cite{scalo86},
   following calculations
  by \cite{schulz02}. V-band luminosity was calculated adopting V
  magnitudes from SIMBAD\footnote{http://simbad.u-strasbg.fr} and a
  common distance modulus of (m-M)~=~18.9 (average value calculated
  from literature by NED
  website\footnote{http://nedwww.ipac.caltech.edu/cgi-bin/nDistance?name=SMC}).
For the clusters HW~1, Lindsay~3 and 113, the V magnitudes were not
found, therefore we did not calculate their masses. 

\begin{center}
\begin{table*}
\caption[Parameters from the literature]{Parameters from the literature.} 
\label{dadosliteratura} 
\begin{tabular}{l@{}l@{}l@{}l@{}l@{}l@{}l@{}}
\hline
  {Cluster} &
  {Metallicity} &
  {Error} &
  {Age (Gyr)} &
  {Error} &
  {Method\,\,} &
  {Reference} \\ 
\hline

 HW 1 
&  -0.7		& --- 		&  4 - 6	& ---		& RP		&  Dias et al., {\it in prep.}  \\

 K3, L8, ESO28SC19  
& -1.5,-0.6,-1.4$^1$& ---	& 3		& ---		& RS, RP	& \cite{1980IAUS...85..305G} \\
& -1.0$^1$	& ---		& 4		& --- 		& RP		& \cite{1981apgc.conf..223G} \\
& $\sim$ -1.2$^2$& ---		& 5 - 8 	& ---		& RP		& \cite{1984ApJ...286..517R} \\
& -1.5$^2$ 	& $\pm$0.2	& $\geq$10 	& ---		& IP		& \protect{\cite{1986A&A...156..261B}} \\
& -1.26$^1$ 	& $\pm$0.10	& 8 - 10 	& ---		& RP		& \cite{1996AJ....112.2004A} \\
& -1.12$^1$, 
    -0.98$^1$ 	& $\pm$0.12, 
		            $\pm$0.12	& ---		& ---		& RS		& \cite{dacosta98} \\
& -1.00$^1$ 	& $\pm$0.28	& 3.5 	& $\pm$1.5		& IS		& \protect{\cite{freitas98}} \\
& -1.16$^1$ 	& $\pm$0.09	& 4.7, 6.0 	& $\pm$0.6, 
						  $\pm$1.3	& RP		& \cite{mighell98} \\
& -1.20$^1$ 	& $\pm$0.2	& 7.0 		& $\pm$1	& IS		& \protect{\cite{piatti05b}} \\
& --- 		& ---		& 6.5 		& $\pm$0.5	& RP		& \cite{glatt08b} \\

 L3, ESO28SC13 
&  ---    	& ---		& 1 - 5		& ---		& RP		& \protect{\cite{kontizas80}} \\
&  ---		& ---		& 0.4		& $\pm$0.1	& RP		& \cite{hodge83} \\ 
&  -0.7		& --- 		& 1 - 2		& ---		& RP		& Dias et al., {\it in prep.} \\

 L11, K7, ESO28SC22 
& ---		& ---		& 1 - 5		& ---		& RP		& \protect{\cite{kontizas80}} \\
& ---		& ---		& 0.3		& $\pm$0.1	& RP		& \cite{hodge83} \\
& -0.93$^2$	& ---		& 3.5		& 1.0		& RP		& \cite{1992ApJS...82..489M} \\
& -0.80$^1$, 
  -0.81$^1$ 	& $\pm$0.14
		  $\pm$0.13	& --- 		& ---		& RS		& \cite{dacosta98} \\
& --- 		& ---		& 3.5 		& $\pm$0.5	& IS		& \protect{\cite{piatti05b}} \\
& -0.6$^2$,
  -0.3$^2$	& ---		& 5.9,	 	
				  4.0		& $^{+1.4}
						  _{-0.8}$,
						  $^{+0.9}
						  _{-0.6}$	& IP		& \cite{2005AJ....129.2701R} \\

 L113, ESO30SC4 
& -1.4$^2$ 	& $\pm$0.2		& 4 - 5		& ---		& RP		& \cite{1984ApJ...280..595M} \\
& -1.44$^1$, 
  -1.17$^1$ 	& $\pm$0.16
		  $\pm$0.12	& --- 		& ---	& RS		& \cite{dacosta98} \\
& -1.24$^1$ 	& $\pm$0.11	& 4.0, 5.3 	& $\pm$0.7, 
						  $\pm$1.3	& RP		& \cite{mighell98} \\

 NGC 121, L10, K2, ESO50SC12
& -1.5,-1.1,-1.4$^1$& ---	& 12		& ---		& RS, RP	& \cite{1980IAUS...85..305G} \\
& ---		& ---		& 13		& $\pm$5	& RP		& \cite{hodge83} \\
& ---		& ---		& 12, 9		& $\pm$2, 
						  $\pm$2	& RP		& Stryker et al. (1985) \\
& -1.3$^2$ 	& $\pm$0.2	& $\geq$10 	& ---		& IP		& \protect{\cite{1986A&A...156..261B}} \\
& -1.46$^1$, 
  -1.19$^1$ 	& $\pm$0.10
		  $\pm$0.12	& --- 		& ---		& RS		& \cite{dacosta98} \\
& -1.20$^1$ 	& $\pm$0.32	& 12 	& $\pm$5		& IS		& \protect{\cite{freitas98}} \\
& -1.71$^1$ 	& $\pm$0.10	& 10.6, 11.9 	& $\pm$0.7,
						  $\pm$1.3	& RP		& \cite{mighell98} \\
& -1.03$^1$ & ---		& 10.6		& $\pm$0.5	& RP		& \cite{dolphin01} \\
& --- 		& ---		& 12 		& $\pm$1	& IS		& \protect{\cite{2002A&A...393..855A}} \\
& --- 		& ---		& 11.8, 11.2,	& $\pm$0.5,	
						  $\pm$0.5,	& RP		& \cite{glatt08a} \\
& --- 		& ---		& 10.5, 10.9,	& $\pm$0.5,
						  $\pm$0.5,	& RP		& \cite{glatt08a} \\
& --- 		& ---		& 11.5, 10.8	& $\pm$0.5,
						  $\pm$1.0,	& RP		& \cite{glatt08a} \\

 NGC 152, L15, K10, ESO28SC24
& ---		& ---		& 1 - 5		& ---		& RP		& \protect{\cite{kontizas80}} \\
& ---		& ---		& 0.61		& $\pm$0.09 	& RP		& \cite{hodge83} \\
& -1.25$^2$ 	& $\pm$0.25	& 3.2 		& $\pm$0.3	& IP		& \protect{\cite{1986A&A...156..261B}} \\

 NGC 222, L24, K19, ESO29SC4, SMC0009(OGLE) 
& -0.6$^2$	& ---		& 0.10	 	& $\pm$0.01	& RP		& \cite{1999AcA....49..157P} \\
& --- 		& ---		& 0.070 	& $\pm$0.007	& RP		& \protect{\cite{2000A&AS..146...57D}} \\
& -0.6$^2$,
  -0.3$^2$	& ---		& 0.092,	 	
				  0.100		& $^{+0.018}
						  _{-0.004}$,
						  $^{+0.046}
						  _{-0.002}$	& IP		& \cite{2005AJ....129.2701R} \\
& -0.3$^2$	& ---		& 0.10 		& $\pm<$0.03	& RP		& \protect{\cite{chiosi06}} \\

 NGC~256, L30, K23 ESO29SC11, SMC0032(OGLE) 
&  -0.6$^2$  	& ---		& 0.10  	& $\pm$0.01	& RP		& \cite{1999AcA....49..157P} \\
&  ---  	& ---		& 0.05    	& $\pm$0.01	& IS		& \protect{\cite{2002A&A...393..855A}} \\
&  -0.3$^2$  	& ---		& 0.10	 	& $\pm<$0.03	& RP		& \protect{\cite{chiosi06}} \\

 NGC~269, L37, K26, ESO29SC16, SMC0046(OGLE) 
& --- 		& ---		& 0.6 		& $\pm$0.2	& IS		& \protect{\cite{piatti05b}} \\
& -0.3$^2$	& ---		& 0.3	 	& $\pm>$0.2	& RP		& \protect{\cite{chiosi06}} \\

 NGC~294, L47, ESO29SC22, SMC0090(OGLE) 
& -0.6$^2$	& ---		& 0.32	 	& $\pm$0.3	& RP		& \cite{1999AcA....49..157P} \\ 
& --- 		& ---		& 0.30 		& $\pm$0.05	& RP		& \protect{\cite{2000A&AS..146...57D}} \\
& --- 		& ---		& 0.3 		& $\pm$0.1	& IS		& \protect{\cite{piatti05b}} \\ 
& -0.6$^2$,
  -0.3$^2$	& ---		& 0.42,	 	
				  0.32		& $^{+0.04}
						  _{-0.08}$,
						  $^{+0.03}
						  _{-0.11}$	& IP		& \cite{2005AJ....129.2701R} \\
& -0.3$^2$	& ---		& 0.4 		& $\pm<$0.2	& RP		& \protect{\cite{chiosi06}} \\

 NGC~361, L67, ESO51SC12
& ---		& ---		& 4		& ---		& RP		& \cite{1981apgc.conf..205H} \\
& ---		& ---		& $>$0.5	& ---		& RP		& \cite{hodge83} \\
& -1.25$^2$ 	& $\pm$0.2	& 8 		& $\pm$1.5	& IP		& \protect{\cite{1986A&A...156..261B}} \\
& -1.45$^1$ 	& $\pm$0.11	& 6.8, 8.1 	& $\pm$0.5,
						  $\pm$1.2	& RP		& \cite{mighell98} \\

 NGC~419, L85, K58, ESO29SC33, SMC0159(OGLE)
& ---		& ---		& 0.67		& $\pm$0.05	& RP		& \cite{hodge83} \\
& -1.2$^2$ 	& $\pm$1.2	& 3.5 		& $\pm$0.3	& IP		& \protect{\cite{1986A&A...156..261B}} \\
& -0.60$^1$ 	& $\pm$0.21	& 1.2 	& $\pm$0.5		& IS		& \protect{\cite{freitas98}} \\
& -0.6$^2$	& ---		& $>$1 		& ---		& RP		& \cite{1999AcA....49..157P} \\
& ---	 	& ---		& 1.2 		& $\pm$0.4	& IS		& \protect{\cite{piatti05b}} \\
& -0.3$^2$	& ---		& 0.4 		& $\pm>$0.2	& RP		& \protect{\cite{chiosi06}} \\
& --- 		& ---		& 1.2 - 1.6	& ---		& RP		& \cite{glatt08b} \\

 NGC~458, L96, K69, ESO51SC26
& ---		& ---		& 0.05		& $\pm$0.01	& RP		& \cite{hodge83} \\
& -0.23	 	& ---		& 0.13 		& $\pm$0.06	& IS		& \protect{\cite{piatti05b}} \\
& -0.6$^2$,
  -0.3$^2$	& ---		& 0.100,	 	
				  0.100		& $^{+0.002}
						  _{-0.028}$,
						  $^{+0.018}
						  _{-0.002}$	& IP		& \cite{2005AJ....129.2701R} \\

\hline
\end{tabular}
\begin{flushleft}{Metallicity values refer 
to 1: [Fe/H], 2: [M/H]=[Z]=[Z/Z$_{\odot}$], assuming Z$_{\odot}$=0.017.
The fields with more than one value refer to distinct criteria adopted by the authors.
The methods are indicated by ``RP'' or ``IP'' for resolved or integrated photometry, and ``RS'' 
or ``IS'' for resolved or integrated spectroscopy.}\end{flushleft}
\end{table*}
\end{center}

\subsection{Full spectrum fitting}
We obtained ages and metallicities for the sample clusters
through the comparison of their integrated spectrum to
SSP models available in literature. 
Modern techniques of spectral 
fitting allow comparison of observations and models on a pixel-by-pixel basis,
and we adopted the public codes {\sc Starlight}
and \emph{ULySS}, described briefly below. 

{\sc Starlight}\footnote{http://www.starlight.ufsc.br} 
\citep{2005MNRAS.358..363C} 
is a multi-purpose code that combines $N$ spectra from a user-defined
base (in our case, SSP models from literature, characterized by age and [Fe/H]) 
in search for the linear
combination which best matches an input observed spectrum.
A {\sc Starlight} run returns the best population mixture that fits the 
observed spectrum, in the form of the light fraction contributed
by each of the SSPs in the base. 
It also returns (i) an estimation of the extinction $A_V$,
(ii) the percentage mean deviation over all fitted pixels 
$\bar{\Delta} = \mid O_{\lambda} - M_{\lambda}\mid / O_{\lambda}$
 (where $O_{\lambda}$ and $M_{\lambda}$ are
the observed and the model spectra, respectively), (iii)
$\mathrm{\chi^2_{red}}$ (reduced chi-square);
of both the global fit, and (iv) the fits to each of the SSPs in the base models.

The use of {\sc Starlight} to study the integrated spectra of clusters
has been extensively discussed in \cite{cid_delgado10}. It is
generally accepted that the majority of stellar clusters can be
represented by a SSP, ideally only one component (SSP model) in the
base would have a non-zero contribution. In practice,
however, a multi-component fit may be returned by the code if (i) the
parameters coverage of the base models is coarse, (ii) there is
contamination from background or foreground field stars, (iii) the S/N is low,
and/or (iv) if any stellar evolution phase present in the population is lacking
in the models (as studied e.g. in \citealp{ocvirk10}).
We adopted as results the mean parameters of the fit (instead of the SSP chi-square selection as in  
Cid Fernandes \& Gonzalez Delgado 2010), as a way to compensate for the  
coarseness of the parameters coverage of the SSP models. The results  
are then given by

\begin{center}
\begin{equation}
\langle({\rm age})\rangle = \sum_{j}{x_j \cdot ({\rm age})_j}
\end{equation}
\end{center}

\noindent and

\begin{center}
\begin{equation}
\langle[Z/Z_{\odot}]\rangle = \sum_{j}{x_j \cdot [Z/Z_{\odot}]_j}
\end{equation}
\end{center}

\noindent where \textbf{x}$_j$ gives the normalized light-fraction of the $j$th SSP
component of the model fit ($\sum_{j}{\textbf{x}_j}=1$).

\emph{ULySS}\footnote{http://ulyss.univ-lyon1.fr} 
\citep{koleva+09}  is a software package
 performing spectral fitting in two astrophysical contexts: 
the determination of stellar atmospheric parameters and the study
of the star formation and chemical enrichment history of galaxies. 
In \emph{ULySS}, an observed spectrum is fitted against a model 
(expressed as a linear combination of components)
through a non-linear least-squares minimization.
In the case of our study, the components are SSP models
(the same as the base models for {\sc Starlight}). 
The grid of SSPs is spline-interpolated to provide a continuous function.
We used the procedures in the package yielding the SSP-equivalent
parameters for a given spectrum, and adopted the values in Table \ref{tab_lit} as
initial guesses. We
noticed that the use of adequate initial
guesses increases the accuracy (and the homogeneity among different
SSP models) of the derived parameters, especially
for metallicities \citep[see discussion in][]{koleva+08}. There
is no equivalent to initial guesses in {\sc Starlight} runs. Also in
contrast with {\sc Starlight}, which fits both the slope of the
spectrum and spectral lines, \emph{ULySS} normalises model and
observation through a multiplicative polinomial in the
model, determined during the fitting process. Therefore, \emph{ULySS}
is not sensitive to flux calibration, galactic extinction, or any other
cause affecting the shape  of the spectrum.

For the present study we adopted three sets of SSP models: 
\begin{itemize}

\item models by \cite{2003MNRAS.344.1000B} (hereafter BC03\footnote{http://www2.iap.fr/users/charlot/bc2003/}), based 
on STELIB stellar library \citep{STELIB} and \cite{bertelli+94} isochrones.
The models cover ages in the range 
$10^5 <$ t(yr) $< 1.5\times10^{10}$, metallicities 0.0001 $<$ Z $<$ 0.05, 
in the wavelength interval 320 - 950 nm (the medium resolution set of models), 
at FWHM $\sim$ 3 {\rm \AA}.

\item models by \cite{2004A&A...425..881L} (hereafter PEGASE-HR\footnote{http://www2.iap.fr/pegase/pegasehr/}), based on 
ELODIE library \citep{ELODIE} and \cite{bertelli+94} isochrones.
The models cover ages 
and metallicities in the range $10^7$ $<$ t(yr) $<$ 1.5x10$^{10}$, 
0.0004  $<$ Z $<$  0.05, and
 wavelength interval of 400 - 680 nm, at FWHM $\sim$ 0.55 {\rm \AA}.

\item preliminary models by \cite{vazdekis+10}\footnote{http://www.iac.es/galeria/vazdekis/vazdekis\_models\_ssp.html}
(in press; see also 
\citealp{2007IAUS..241..133V}), which are an extension of the models 
by \cite{1999ApJ...513..224V} using the MILES library \citep{MILES} and 
isochrones by \cite{2000A&AS..141..371G}.
The models cover ages  $10^8$ $<$ t(yr) $<$ 1.5x10$^{10}$, 
metallicities 0.0004  $<$ Z $<$  0.03, and
wavelength interval 350 -740 nm, at FWHM $\sim$ 2.3 {\rm \AA}.

\end{itemize}

By fitting the data with three different sets of SSP models, we aim at 
a better handle on the uncertainties of the derived parameters, and 
possibly detect model dependencies.

\section{Results and discussion}

We reported in Tables \ref{tab_starlight} and \ref{tab_ulyss} the 
ages and metallicities
obtained from {\sc Starlight} and \emph{ULySS} fits, respectively. 
Individual fits are shown in Figs. \ref{fig_st_first} 
to \ref{fig_uly_last} (for the sake of space, only fits with
PEGASE-HR are presented).

\begin{figure*}
\centering
\includegraphics[width=\textwidth]{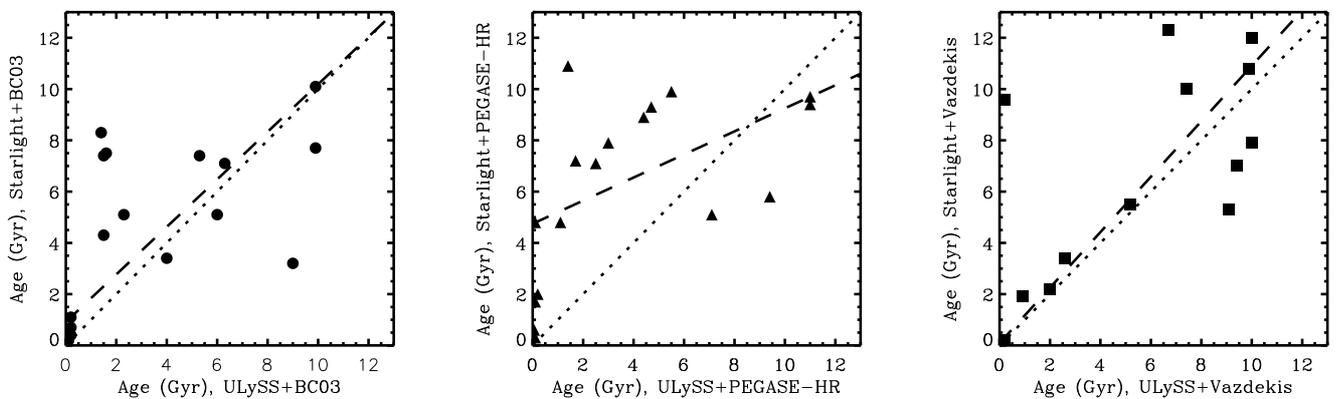}
\caption{Ages derived with {\sc Starlight} and \emph{ULySS} (Gyr)
in each panel using a different SSP model: {\it Left panel:} BC03,
{\it Middle:} PEGASE-HR, {\it Right panel:} Vazdekis et al. The dashed lines are linear fits
to the points using a least absolute deviation method and the dotted
lines are the one-to-one match.}
\label{agecomp}
\end{figure*}

\begin{figure*}
\centering
\includegraphics[width=\textwidth]{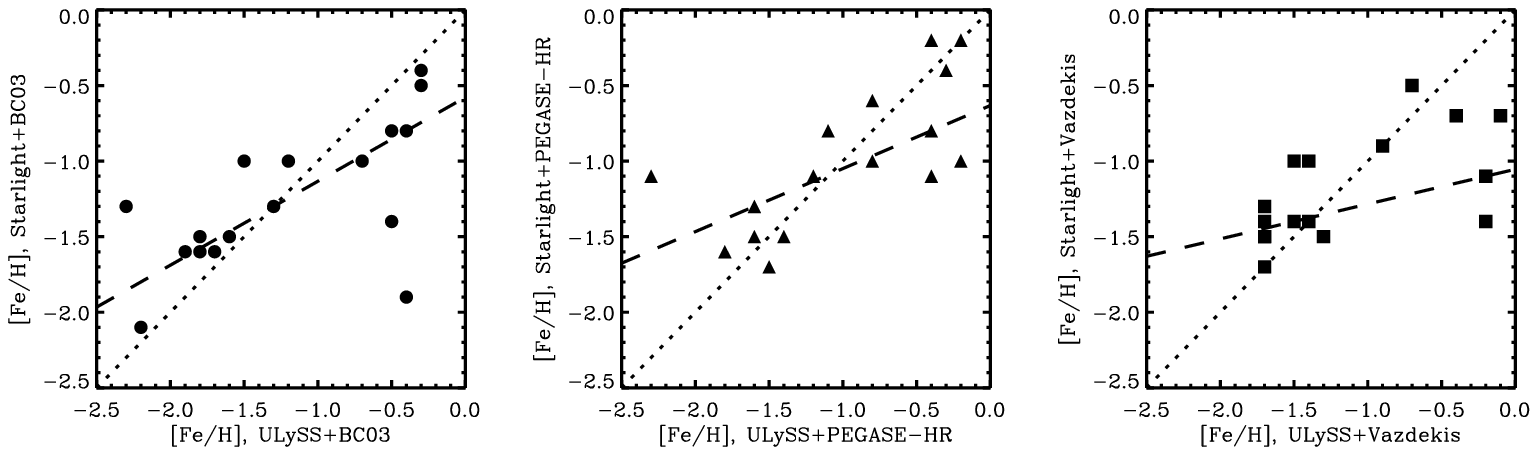}
\caption{Same as Fig. \ref{agecomp} for metallicities.}
\label{metcomp}
\end{figure*}

In Figs. \ref{agecomp} and \ref{metcomp} we compare the results of the
two codes. Figure \ref{agecomp} shows the ages obtained with
\emph{ULySS} in the abcissa and those from {\sc Starlight} in
ordinates, for the BC03, PEGASE-HR, and Vazdekis models, respectively.
We use the least absolute deviation method for the linear fits because
it is considered more robust than $\chi^2$ minimisation.
Figure \ref{metcomp} shows the same plots for metallicities.
In Fig. \ref{agecomp} we note that, with the exception of the  
runs with PEGASE-HR models, ages derived from the two codes show a  
high dispersion, but not a clear trend. In contrast, in the case of  
metallicities (Fig. \ref{metcomp}), Starlight runs will result in higher  
metalliticies than ULYSS runs in the metallicity tail, with the trend  
reversing at the high-metallicity tail. In both cases the dispersion
is high, and a safe conclusion can only be reached with a larger sample.

\begin{figure*}
\centering
\includegraphics[width=\textwidth]{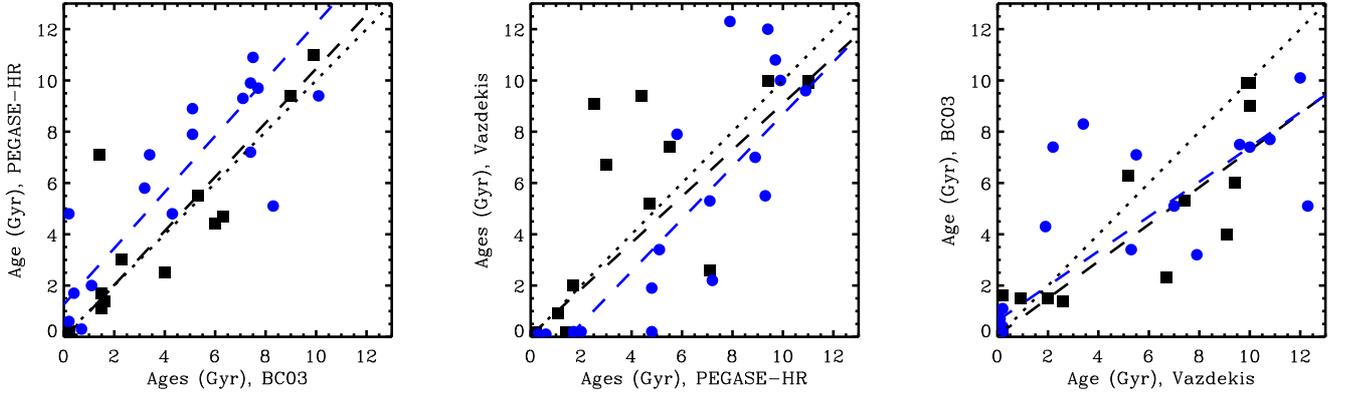}
\caption{Comparisons of ages derived with the three sets of SSPs
  (BC03, PEGASE-HR, Vazdekis). {\sc STARLIGHT} values are shown as
  blue circles, \emph{ULySS} values as black squares. The dashed lines
  are linear fits to the points using a least absolute deviation
  method and the dotted lines are the one-to-one match.} 
\label{setsagecomp}
\end{figure*}

\begin{figure*}
\centering
\includegraphics[width=\textwidth]{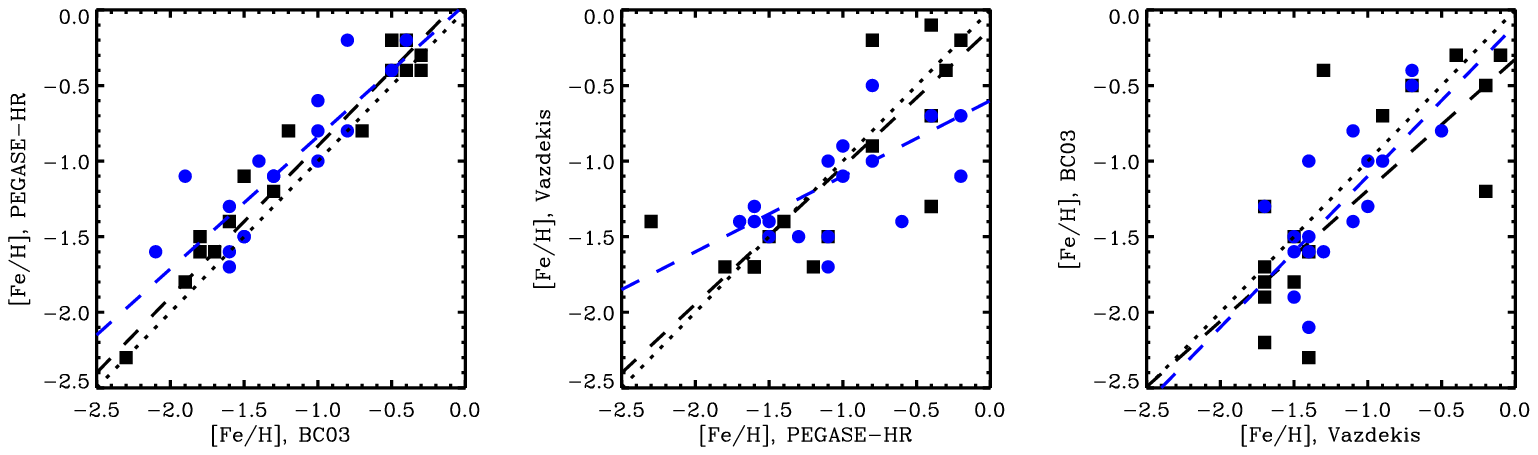}
\caption{Same as Fig. \ref{setsmetcomp} for metallicities.} 
\label{setsmetcomp}
\end{figure*}

Figures \ref{setsagecomp} and \ref{setsmetcomp} compare the results
between different models but the same code, for ages and metallicities,
respectively. With the exception of the middle panel in
Fig. \ref{setsmetcomp}, we see that differences are dominated by
shifts, rather than showing a dependence with age and/or
metallicity. Figures \ref{agecomp} to \ref{setsmetcomp} seem to indicate
that, while the use of different SSP models might introduce zero-point
shifts in the derived parameters, the choice of fitting code might
introduce a more complicated behaviour, dependent on the range of
population parameters studied.

\begin{figure*}
\centering
\begin{minipage}[b]{0.40\textwidth}
\includegraphics[width=\columnwidth]{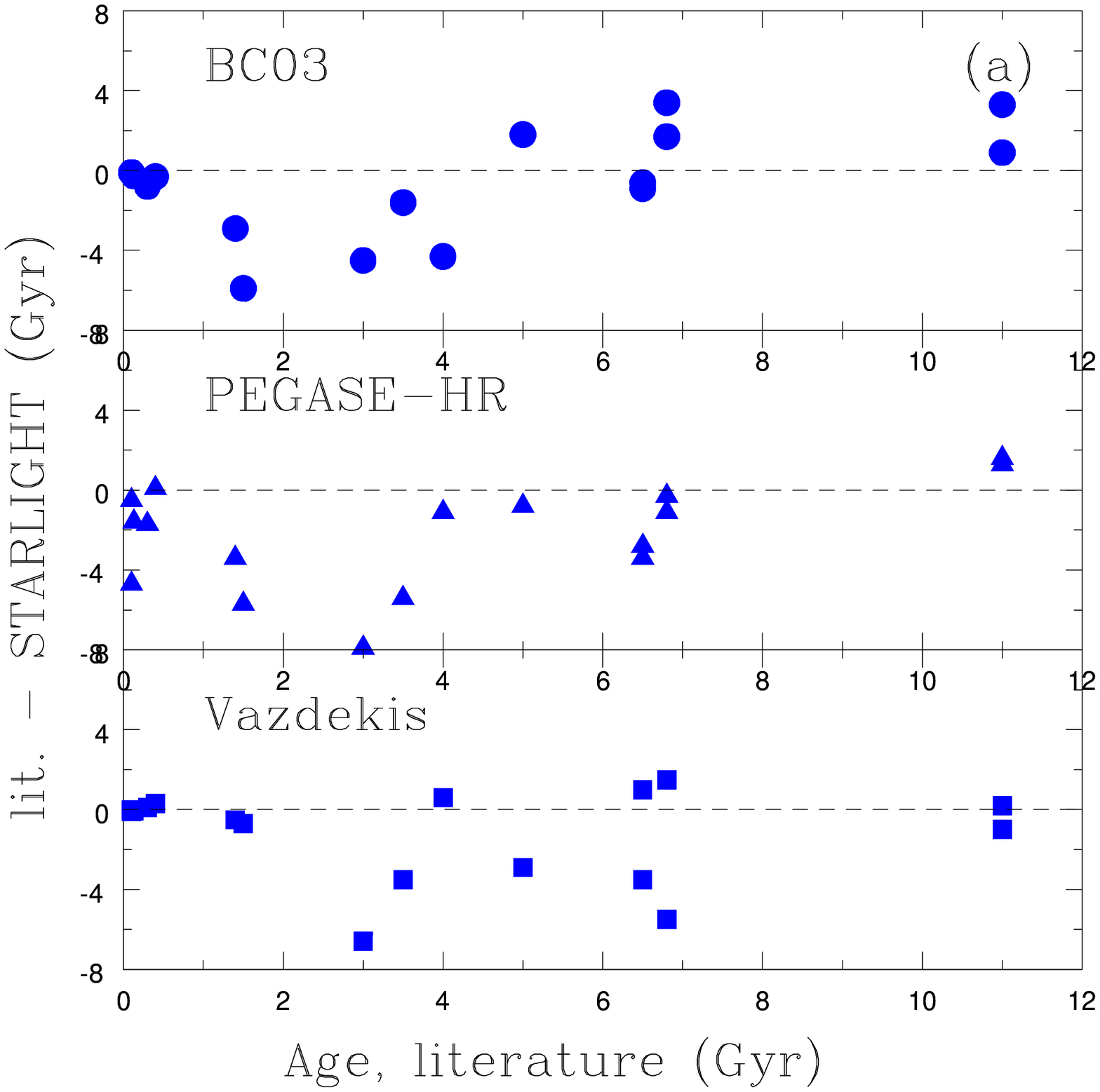}
\end{minipage}
\begin{minipage}[b]{0.40\textwidth}
\includegraphics[width=\columnwidth]{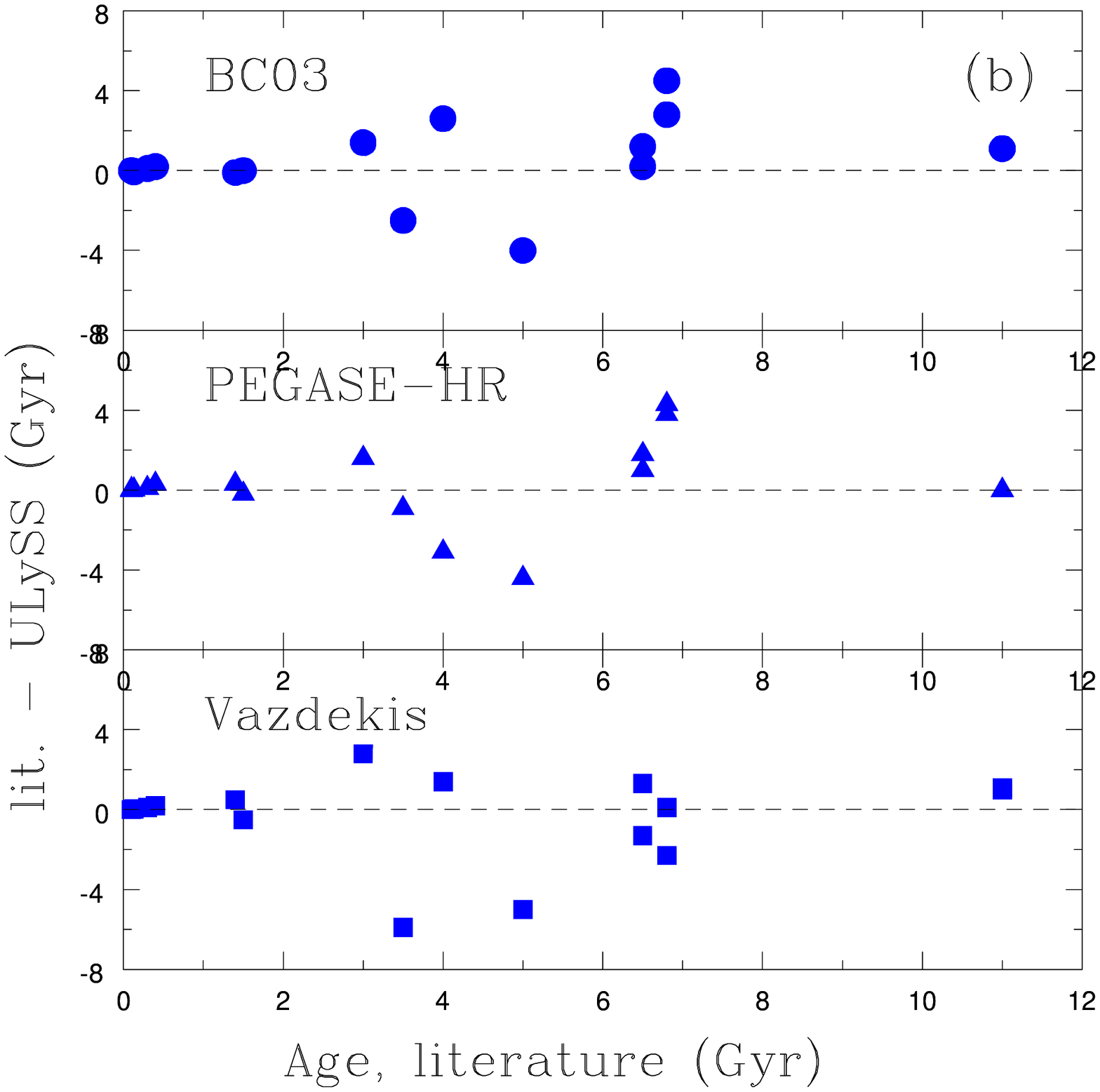}
\end{minipage}
\caption{Ages from the literature (given in log age(yr)) 
 from Table \ref{tab_lit} in the abscissae vs.
difference of age from literature and the result from
{\sc Starlight} and  \emph{ULySS} with the 3 SSPs,
 in the ordinate.
}
\label{agelit}
\end{figure*}

\begin{figure*}
\centering
\begin{minipage}[b]{0.40\textwidth}
\includegraphics[width=\columnwidth]{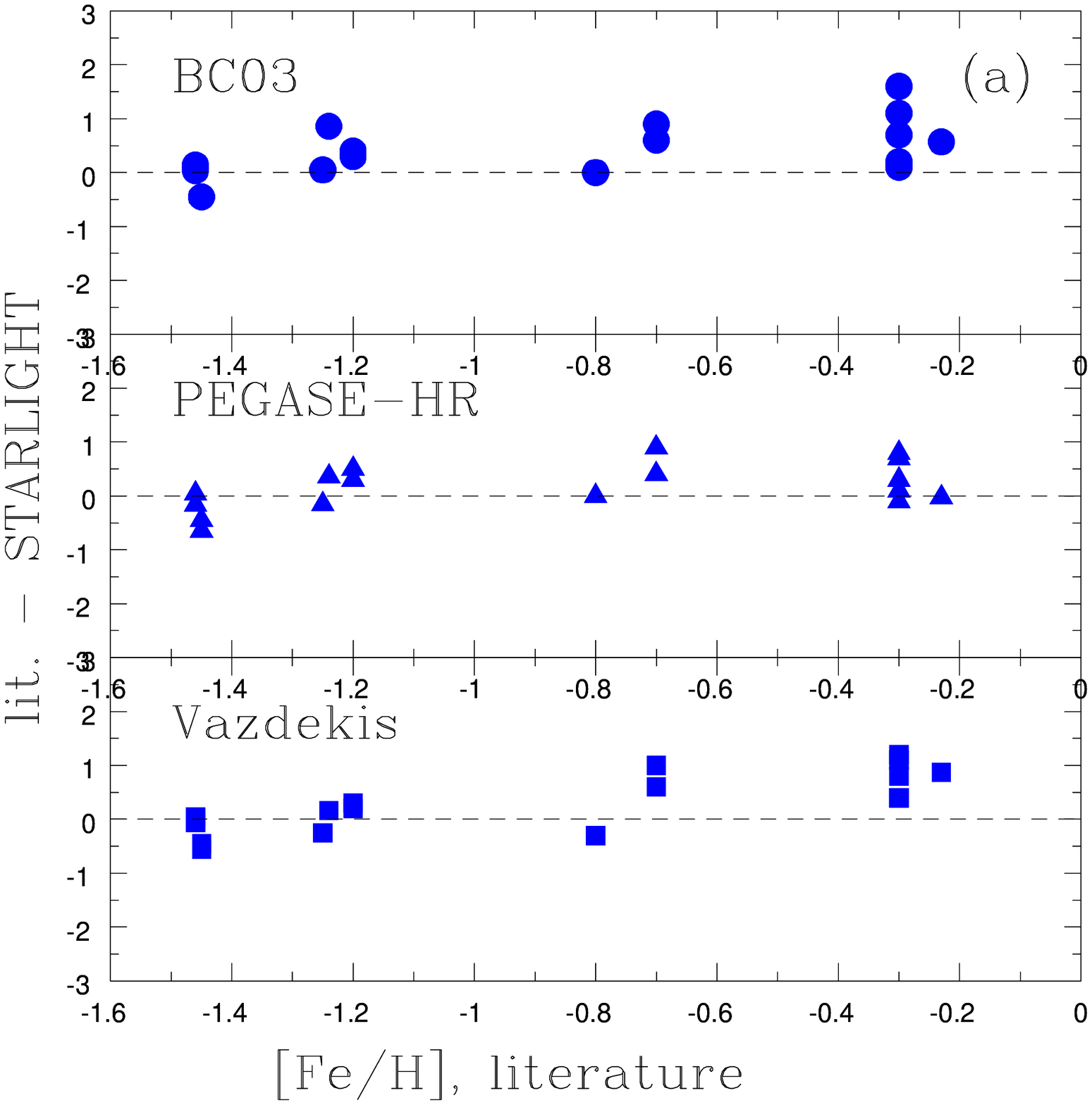}
\end{minipage}
\begin{minipage}[b]{0.40\textwidth}
\includegraphics[width=\columnwidth]{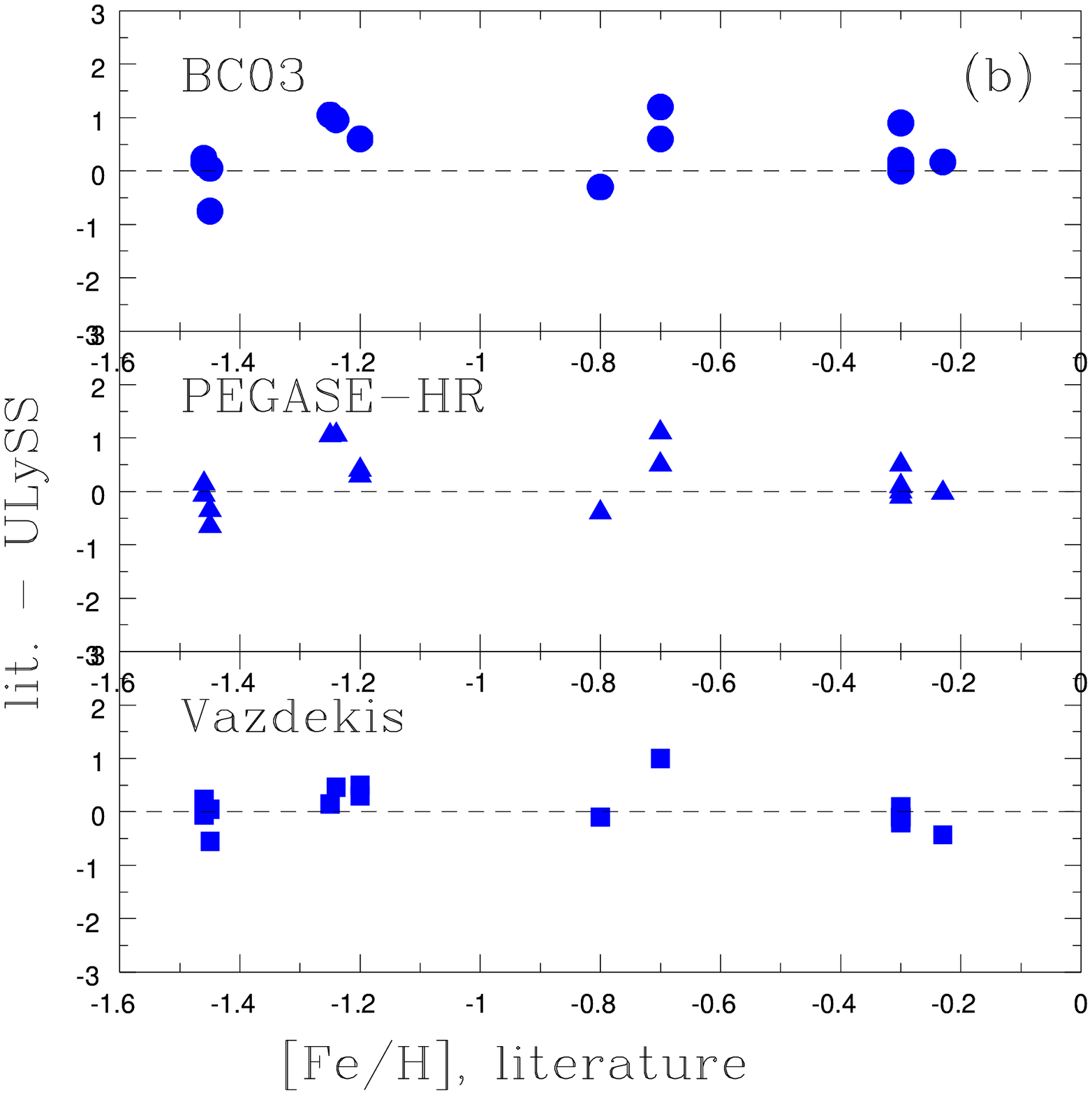}
\end{minipage}
\caption{Same as Fig. \ref{agelit} for metallicities.}
\label{metlit}
\end{figure*}

\begin{figure}
\centering
\includegraphics[width=\columnwidth]{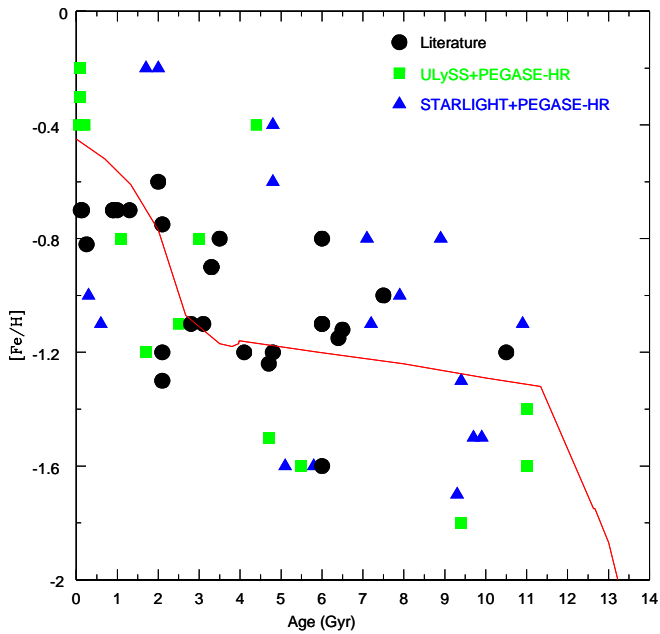}
\caption{Age-metallicity data for the sample clusters and selected
  literature data for well-known clusters.  Symbols: filled circles:
  literature data (Table \ref{lit_known}); filled triangles: present
  results based on {\sc Starlight}+PEGASE-HR; filled triangles:
  present results based on \emph{ULySS}+PEGASE-HR. 
The chemical evolution model by 
\cite{1998MNRAS.299..535P}
is overplotted.}
\label{fig_agemet}
\end{figure}

In Figs. \ref{agelit} and \ref{metlit} we show the comparison of our results
with literature data (Table \ref{dadosliteratura}) for ages and
metallicities, respectively.
Literature data as reported in Table \ref{tab_lit} are plotted vs.
difference of age between literature and the result from {\sc
  Starlight} (panel a) and \emph{ULySS} (panel b) (given in Gyr). 
Figure \ref{metlit} shows the same for metallicity.
In Table \ref{lit_known} we report a list of reliable ages and metallicities
for well-known and/or well studied clusters, by trying to select mostly 
intermediate/old age ones. The list of clusters is basically that of
\cite{carrera08}, \cite{freitas98}, \cite{glatt08a}, \cite{glatt08b},
\cite{bica08}, \cite{glatt09}, and \cite{parisi09}.
Figure \ref{fig_agemet}
 gives the age and metallicity of literature data for well-studied clusters,
as reported in Table \ref{lit_known},  and the results for
our sample clusters derived with \emph{ULySS}+PEGASE-HR. 
As discussed
for example in Da Costa \& Hatzidimitriou (1998), we see 
a relatively rapid rise in metallicity in the first 3 to 5 Gyrs
(assuming that chemical evolution started at 15 Gyr), and a slow
increase in the metallicity from [Fe/H]$\approx$-1.1 to -1.3 to the present
value of [Fe/H]$\approx$-0.7.
 We also overplot the chemical
evolution model for the SMC computed by \cite{1998MNRAS.299..535P}.
 The model fits the confirmed literature data well, as also found in
 previous work, and our results are compatible with the model and
 literature data.

\begin{table}[!htb]
\centering
\caption{Values adopted from the literature (Table
  \ref{dadosliteratura}) for the clusters studied in this work, where masses
   are estimated based on models by \cite{schulz02}. }
\label{tab_lit}
\begin{tabular}{lcccc}
\hline
\noalign{\smallskip}
Cluster &  Age  & [Z/Z$_{\odot}$]  & log(M/M$_{\odot}$) & log(M/M$_{\odot}$) \\
\noalign{\smallskip}
        &  (Gyr) &                 & Salpeter & Scalo \\
\noalign{\smallskip}
\hline
\noalign{\smallskip}
HW1                 & 5      $\pm$  1.0  	& -0.7   $\pm$ ---  & --- & --- \\
K3                    & 6.5   $\pm$  0.5  	& -1.2   $\pm$ 0.2   & 5.0  & 4.6  \\
L3             	& 1.5   $\pm$  0.5  	& -0.7   $\pm$ ---  & --- & --- \\
L11          	& 3.5   $\pm$  0.5  	& -0.8   $\pm$ 0.14 & 4.0  & 3.7  \\
L113                & 4.0   $\pm$  0.7           & -1.24 $\pm$ 0.11 & --- & --- \\
NGC 121 	  	& 11    $\pm$  0.5  	& -1.46 $\pm$ 0.10 & 5.5  & 5.2  \\
NGC 152 		& 3.0   $\pm$  1.0  	& -1.25 $\pm$ 0.25 & 4.4  & 3.9  \\
NGC 222        	& 0.10 $\pm$  0.03	& -0.3   $\pm$ ---  & 3.8  & 3.8  \\
NGC~256        	& 0.10 $\pm$  0.03	& -0.3   $\pm$ ---  & 3.8  & 3.8  \\
NGC~269        	& 0.3   $\pm$  0.2  	& -0.3   $\pm$ ---  & 4.0  & 3.8  \\
NGC~294        	& 0.4   $\pm$  0.2  	& -0.3   $\pm$ ---  & 4.4  & 4.0  \\
NGC~361 		& 6.8   $\pm$  0.5  	& -1.45 $\pm$ 0.11 & 5.1  & 4.7  \\
NGC~419        	& 1.4   $\pm$  0.2  	& -0.3   $\pm$ ---  & 5.3  & 4.9  \\
NGC~458        	& 0.13 $\pm$  0.06	& -0.23 $\pm$ ---  & 3.9  & 4.2  \\
\noalign{\smallskip}
\hline
\end{tabular}
\end{table}

\begin{table*}
\centering
\caption{Best-fit results using {\sc Starlight}.}
\label{tab_starlight}
\begin{tabular}{l|cc|cc|cc}
\noalign{\smallskip}
\hline
\noalign{\smallskip}
 & \multicolumn{2}{c}{BC03} & \multicolumn{2}{c}{PEGASE-HR} & \multicolumn{2}{c}{Vazdekis et al.} \\
\noalign{\smallskip}
\hline
\noalign{\smallskip}
Cluster & Age (Gyr) & [Z/Z$_{\odot}$] & Age (Gyr) & [Z/Z$_{\odot}$] & Age (Gyr) &  [Z/Z$_{\odot}$]  \\
\noalign{\smallskip}
\hline
\noalign{\smallskip}
HW1		&   3.2 &  -1.6 &    5.8 &  -1.6 &     7.9 &  -1.3\\
K3 (LNA)	&   7.1 &  -1.6 &    9.3 &  -1.7 &     5.5 &  -1.4 \\
K3 (ESO) 	&   7.4 &  -1.5 &    9.9 &  -1.5 &    10.0 &  -1.5 \\
L3         	&   7.4 &  -1.3 &    7.2 &  -1.1 &     2.2 &  -1.7 \\
L11         	&   5.1 &  -0.8 &    8.9 &  -0.8 &     7.0 &  -0.5 \\
L113            &   8.3 &  -2.1 &    5.1 &  -1.6 &     3.4 &  -1.4\\ 
NGC 121 (LNA)  	&   7.7 &  -1.5 &    9.7 &  -1.5 &    10.8 &  -1.4 \\
NGC 121 (ESO) 	&  10.1 &  -1.6 &    9.4 &  -1.3 &    12.0 &  -1.5 \\
NGC 152    	&   7.5 &  -1.3 &   10.9 &  -1.1 &     9.6 &  -1.0 \\
NGC 222        	&   0.2 &  -1.9 &    0.6 &  -1.1 &     0.1 &  -1.5 \\
NGC~256        	&   0.2 &  -0.5 &    4.8 &  -0.4 &     0.2 &  -0.7 \\
NGC~269         &   1.1 &  -0.4 &    2.0 &  -0.2 &     0.2 &  -0.7 \\
NGC~294        	&   0.7 &  -1.4 &    0.3 &  -1.0 &     0.1 &  -1.1 \\
NGC~361 (ESO99) &   5.1 &  -1.0 &    7.9 &  -1.0 &    12.3 &  -0.9 \\
NGC~361 (ESO00) &   3.4 &  -1.0 &    7.1 &  -0.8 &     5.3 &  -1.0 \\
NGC~419        	&   4.3 &  -1.0 &    4.8 &  -0.6 &     1.9 &  -1.4 \\
NGC~458        	&   0.4 &  -0.8 &    1.7 &  -0.2 &     0.2 &  -1.1 \\
\noalign{\smallskip} 
\hline
\end{tabular}
\end{table*}

\begin{table*}
\centering
\caption{Best-fit results using \emph{ULySS}.}
\label{tab_ulyss}
\begin{tabular}{l|cc|cc|cc}
\noalign{\smallskip}
\hline
\noalign{\smallskip}
 & \multicolumn{2}{c}{BC03} & \multicolumn{2}{c}{PEGASE-HR} & \multicolumn{2}{c}{Vazdekis et al.} \\
\noalign{\smallskip}
\hline
\noalign{\smallskip}
Cluster 	&  Age (Gyr) & [Z/Z$_{\odot}$] & Age (Gyr) & [Z/Z$_{\odot}$] & Age (Gyr) & [Z/Z$_{\odot}$]  \\
\noalign{\smallskip}
\hline
\noalign{\smallskip}
HW1		&  9.0   & -1.9  & 9.4   & -1.8  &  10   & -1.7 \\
K3 (LNA) 	&  6.3   & -1.8  & 4.7	 & -1.5  &  5.2  & -1.5 \\
K3 (ESO)  	&  5.3   & -1.8  & 5.5   & -1.6  &  7.4  & -1.7 \\
L3         	&  1.5   & -1.3  & 1.7   & -1.2  &  2.0  & -1.7 \\
L11         	&  6.0   & -0.5  & 4.4   & -0.4  &  9.4  & -0.7 \\
L113        	&  1.4   & -2.2  & 7.1   & -2.6  &  2.6  & -1.7 \\ 
NGC~121 (LNA)	&  9.9   & -1.6  & 11	 & -1.4  &  9.9  & -1.4 \\
NGC~121 (ESO)	&  9.9   & -1.7  & 11	 & -1.6  &  10   & -1.7 \\
NGC~152     	&  1.6   & -2.3  & 1.4   & -2.3  &  0.2  & -1.4 \\
NGC~222        	&  0.1   & -0.4  & 0.06  & -0.4  & 0.1   & -1.3 \\
NGC~256        	&  0.1   & -0.3  & 0.1   & -0.3  & 0.1   & -0.4 \\
NGC~269        	&  0.2   & -0.3  & 0.2   & -0.4  & 0.2   & -0.1 \\
NGC~294        	&  0.2   & -0.5  & 0.1   & -0.2  & 0.2   & -0.2 \\
NGC~361 (ESO99)	&  2.3   & -0.7  & 3.0   & -0.8  & 6.7   & -0.9 \\
NGC~361 (ESO00)	&  4.0   & -1.5  & 2.5   & -1.1  & 9.1   & -1.5 \\
NGC~419        	&  1.5   & -1.2  & 1.1   & -0.8  & 0.9   & -0.2 \\
NGC~458        	&  0.2   & -0.4  & 0.1   & -0.2  & 0.1   &  0.2 \\
\noalign{\smallskip} 
\hline
\end{tabular}
\end{table*}

\begin{table}
\centering
\caption{Literature data of age and metallicity for well-known
SMC star clusters.
}
\label{lit_known}
\begin{tabular}{lc@{}c@{}c@{}|l@{}c@{}c@{}c}
\noalign{\smallskip}
\hline
\noalign{\smallskip}
Cluster & \phantom{-}[Fe/H] & \phantom{-}Age (Gyr)  & \phantom{-}ref.\, &
 \phantom{-}Cluster & \phantom{-}[Fe/H] & \phantom{-}Age (Gyr) & \phantom{-}ref. \\
\noalign{\smallskip}
\hline
\noalign{\smallskip}
\hline
\noalign{\smallskip}
BS 90  & -1.0 & 4.3  & 1  &  L13          &  -1.24 & 4.7  &  9    \\
BS 121 	& -1.2 	& 2.3	& 2,3   &  L19 	        & -0.75 & 2.1 	& 2,3   \\
BS 196  & -1.7 & 5.0  & 4  &  L27 	        & -1.3 	& 2.1 	& 2,3   \\ 
HW 47 	& -1.0 	& 2.8 	& 2,3   &  L106         & -0.7 	& 0.9  	& 10    \\
HW 84 	& -1.2 	& 2.4 	& 2,3   &  L108         & -0.7 	& 0.9  	& 10    \\
HW 86 	& -0.75 & 1.6 	& 2,3   &  L 110   & -1.15 & 6.4   & 3     \\
K28 	& -1.2 	& 2.1 	& 5     &  L111 	& -0.7 	& 1   	& 10    \\
K44 	& -1.1 	& 3.1 	& 5     &  L113 	& -1.1 & 6  	& 8     \\
K3 	& -1.12 & 6.5  	& 6     &  L114 	& -0.7 	& 0.14 	& 10    \\
L32 	& -1.2 	& 4.8  	& 5     &  L115 	& -0.7 	& 0.11 	& 10    \\
L38 	& -1.65 & 6.0 	& 5     &  L116 	& -1.1 	& 2.8 	& 5     \\
L1 	& -1.0 & 7.5 	& 7     &  NGC121 	& -1.2 & 10.5  	& 7,8   \\
L 4    & -0.9  & 3.3   & 2,3    &  NGC330 	& -0.82  & 0.25	& 11    \\ 
L5 	& -1.2 	& 4.1 	& 2,3   &  NGC339 	& -1.1   & 6.0  & 7     \\
L6 	& -0.9 	& 3.3 	& 2,3   & NGC 411 & -0.7   & 1.3   & 12   \\
L7 	& -0.6 	& 2.0 	& 2,3   & NGC416 	& -0.8 & 6.0 & 7,12 \\
L11 	& -0.8 	& 3.5  	& 8     &               &        &       &       \\
\noalign{\smallskip}
\hline 
\end{tabular}
\begin{flushleft}
{References:
1 \cite{sabbi07};
2 \cite{piatti05a}, \cite{piatti05b};
3 \cite{parisi09}
4 \cite{bica08}; 
5 \cite{piatti01};
6 \cite{glatt08b};
7 \cite{glatt09}; 
8 \cite{dacosta98};
9 \cite{mighell98}; 
10 \cite{piatti07}; 
11 \cite{hill99}; 
12 \cite{freitas98}.}
\end{flushleft}
\end{table}

\section{Comments on individual clusters}

 Results obtained with {\sc Starlight}
and \emph{ULySS}, given in Tables \ref{tab_starlight} and
\ref{tab_ulyss}, for each cluster, are compared with previous analyses.

\subsection{HW 1} 
There are no literature data on this cluster. From the analysis carried out
with {\sc Starlight}, we get an intermediate/old age and a low metallicity,
whereas with \emph{ULySS}, we get an old age and a low metallicity,
and with both codes the results are consistent
among the three sets of SSPs. The identification of such
an old and metal-poor cluster is an important result. Preliminary
CMD data obtained in our group indicate an age 
around 6 Gyr (to be published elsewhere), in better
agreement with {\sc Starlight} results.

\subsection{Kron 3} 
\cite{glatt08a} derived an age of 6.5 Gyr, whereas Rich et al. (1984)
find an age of 5-8 Gyr.
The ages inferred are in most cases similar or older than the 6.5 Gyr expected.
Low metallicities around [Fe/H]$\approx$-1.6 are retrieved in all runs.
High-resolution spectroscopy of individual stars of this cluster would be
of great interest.

\subsection{Lindsay 3} 
\cite{kontizas80} subdivided a sample of 20 star clusters 
in young or old based on the colour of the nuclei of each
cluster. By using this method, Lindsay~3 is 1 to 5~Gyr old,
therefore an intermediate/old age cluster.
Ages both from {\sc Starlight} and \emph{ULySS}
give either around 1.5 or 7 Gyr. This cluster has among the lowest S/N in our
sample, so this age discrepancy is not surprising. (Cid Fernandes \&
Gonzales Delgado 2010 suggest a minimum of S/N~$\sim$~30 to obtain robust
results.) Preliminary CMD analyses
in our group, to be published elsewhere, give
an age of 1-2 Gyr and metallicity of [Fe/H]$\approx$-0.7.
The {\sc Starlight} and \emph{ULySS} metallicities all agree that the cluster is
metal-poor, likewise the CMD indications.
 L3 is revealed as an intermediate/old age and low metallicity cluster.

\subsection{Lindsay 11} 
\cite{kontizas80} gives and age in the range 1 to 5 Gyr based
on CMDs. Both codes and the three SSPs give
moderate metallicities within -0.8$<$[Fe/H]$<$-0.5,
in good agreement with \cite{dacosta98}.
Ages of 5.1 to 8.9 Gyr are found with {\sc Starlight}, 
whereas \emph{ULySS} gives ages between 4.4 and 9.4.
 This could be an interesting cluster
with an intermediate/old age around 5 Gyr.

\subsection{Lindsay 113} 
\cite{mighell98} have derived and age of 4.7~Gyr
(in the range of 4.0 $<$ t(Gyr) $<$ 5.3) and [Fe/H]=-1.24.
{\sc Starlight} and \emph{ULySS}  give low metallicities.
The ages from {\sc Starlight} are similar or older
(for BC03) than \cite{mighell98} value, whereas a higher age
dispersion is found with \emph{ULySS}. This cluster of
intermediate/old age is very promising and should be observed
further.

\subsection{NGC 121}
\cite{glatt08a} obtained an HST/ACS CMD of NGC 121
and derived ages of 11.8, 11.2, and 10.5 Gyr based on Teramo 
\citep{2004ApJ...612..168P},
 Padova \citep{2000A&AS..141..371G} and Dartmouth \citep{2007ApJ...666..403D}
isochrones. In a final age scale, the authors adopt an age between
10.9 and 11.5$\pm$0.5 Gyr.
From our runs, metallicities in the range  -1.7  $<$  [Fe/H] $<$ -1.3 are obtained.
All ages are in the range 7.7 $<$ t(Gyr) $<$ 12.
 This well-known oldest cluster of the SMC could be a survivor of an epoch
of a somewhat  delayed first burst of cluster formation
 in the SMC (\citealp{glatt08a}).

\subsection{NGC 152} 
For this cluster an age range of 1 to 5 Gyr is also given by
\cite{kontizas80}. {\sc Starlight} gives old ages of 7.5 to 10.9 Gyr
and metallicities around [Fe/H]$\approx$-1.1.
\emph{ULySS} gives young ages of 0.2 to 1.5 Gyr
and very low metallicities of -2.3$<$[Fe/H]$<$-1.4.
For this cluster both
the age and metallicity remain undefined, and it is
clearly a good candidate for further studies on intermediate/old
age clusters.

\subsection{NGC 222}

A young age of 100 Myr and [Fe/H]=-0.3 for NGC 222,
were derived by employing isochrone fitting to VI CMDs by \cite{chiosi06}. 
Young ages are obtained in all cases. On the other hand,
low metallicities are derived in most cases and it
  remains undefined.

\subsection{NGC 256}

\cite{chiosi06} give 100 Myr and [Fe/H]=-0.3.
{\sc Starlight} gives similar results using BC03 and Vazdekis et al.,
whereas PEGASE-HR gives an intermediate age, with a metallicity
  similar to that derived by \cite{chiosi06}.
 \emph{ULySS} gives ages and metallicities in agreement with Chiosi et al. with the
   three SSP sets.

\subsection{NGC 269}

\cite{chiosi06} report 300 Myr and [Fe/H]=-0.3.
\emph{ULySS} gives results similar to Chiosi et al. values, with the 3 SSPs.
{\sc Starlight} gives metallicities -0.7~$<$~[Fe/H]~$<$~-0.2, however
BC03 and PEGASE-HR give intermediate ages.

\subsection{NGC 294}  

Pietrzy\'nski \& Udalski (1999) present the CMD of NGC 294,
and have derived an age of 0.33$\pm$0.3 Gyr and a metallicity of [Fe/H]$\approx$-0.6.
The young age is confirmed in all runs.
\emph{ULySS} metallicities agree with the literature value, whereas {\sc
  Starlight} runs give lower metallicities around [Fe/H]$\approx$-1.2.

\subsection{NGC 361} 

The NGC~361 CMD from \cite{mighell98} was
cleaned of contaminations by field stars. There are
 two predominant populations: one 
older than the cluster that has similar CMD components, 
and the other one younger that has an extended main sequence.
The cleaned cluster  CMD shows clear RGB and HB
sequences. Metallicities were derived from 
the CMDs using two methods and combining the results. The first
method was the simultaneous fit of reddening and metallicity method
(\citealp{1994AJ....107..618S}), that depends on the magnitude level 
of the HB, the colour of the RGB at the level of the HB, and
the shape and position of the RGB. The second one was the
RGB slope method. A metallicity of
[Fe/H]=-1.45~$\pm$~0.11 was adopted.
The method to determine the age was based on
the colour of the red HB and the RGB at the level of the HB
(\citealp{1995ApJ...450..712S}) for a given metallicity,
and an age of 6.8~$\pm$~0.5~Gyr was adopted. They tried another method
to derive relative ages with respect to Lindsay 1 and found
8.1~$\pm$~1.2~Gyr.
A population like this is clearly not well modelled by SSPs,
  and a challenge for spectral analysis such as the one presented here. 
Indeed an inspection of the multi-population fits returned by  
STARLIGHT show superpositions of old ($\sim$~10~Gyr) and intermediate-age  
($\sim$~2~Gyr) populations for this cluster.
  Nevertheless, the literature
  value of metallicity around $\sim$~-1.3 (see also Table
  \ref{dadosliteratura}) is confirmed in all combinations of code and
  SSPs.
Ages are retrieved in a wide range, between 2.3 to 12.3~Gyr. 
{\sc Starlight}+PEGASE-HR and \emph{ULySS}+Vazdekis are the fits that
better match Mighell et al. (1998) age results.

\subsection{NGC 419} 

Using HST/ACS data, \cite{glatt08b} have recently 
demonstrated that NGC 419 is among the most interesting populous
stellar clusters in the SMC due to the clear presence 
of multiple stellar populations with ages between $\sim$ 1.0 
and $\sim 2.0$ Gyr.
This hypothesis was confirmed by a detailed analysis of this
HST/ACS CMD performed by \cite{girardi09}, which sustained
the presence of multiple stellar populations not only
by the main sequence spread, but also by a clear presence of
a secondary clump. Furthermore, very recently
Rubele, Girardi \& Kerber (2009, in prep.) have recovered the SFH for
this cluster, which lasts at least 700 Myr 
with a marked peak at the middle of this interval, for an
age of 1.5 Gyr. Assuming the same chemical composition
for all stars in NGC 419, these authors also determined a
metallicity of [Fe/H]=-0.86 $\pm$ 0.09.   

The same caveats of the previous cluster, on trying to fit SSPs
  to such a complex population, applies for this cluster as well. Even
  so, most combinations give satisfactory results, compatible with 
\cite{glatt08b}. The older ages retrieved by {\sc Starlight}+BC03, +PEGASE-HR
could be due
to the double turn-off found by \cite{glatt08b},
where isochrones from 1 to 3 Gyr were fitted.
\emph{ULySS} gives ages and metallicities in agreement with Glatt et al. (2008b).

\subsection{NGC 458} 

From integrated spectroscopy, \cite{piatti05a} give
130 Myr, and [Fe/H]=-0.23.
Young ages are derived in all fits with the exception of
{\sc Starlight}+PEGASE-HR.
Metallicity values show a rather large dispersion, confirming
that uncertainties on this parameter are larger for young ages.

\section{Conclusions}
\label{sec_conclusions}

We observed mid-resolution integrated spectra of SMC star clusters to
study the SMC chemical evolution and in particular to determine the
stellar population parameters of intermediate/old age clusters. To
study these integrated spectra, we exploited  the ability of the codes
{\sc Starlight} and \emph{ULySS}, coupled with Single Stellar
Populations (SSPs) spectral models to derive their ages and
metallicities. The SSPs models employed are those by BC03, Pegase-HR,
and Vazdekis et al.

We highlight the importance of the intermediate/old age
clusters HW1, L3, L11, NGC 152, NGC 361, NGC 419, and L113. We also
confirm the intermediate/old age of Kron 3 and old age of NGC 121.

There seems to be an indication that the choice of the code will have more
impact on the results than the choice of models. We point out that the
STARLIGHT results adopted are a mean of the main stellar populations
identified, therefore some of the differences in the results relative
to ULySS may result from  this.

We also derived masses for the sample clusters, reported in Table \ref{tab_lit}.
De Grijs \& Goodwin (2008) published cluster mass functions based on
statistically complete SMC cluster samples, and our results 
 are compatible with their mass distribution  (their Figure 2, panel d), 
since we find that most  clusters have masses around 
log(M$_{cl}$/M$_{\odot}$)~$\sim$~4.

Another interesting issue is the existence of
 very metal-poor stellar populations in the SMC.
Planetary nebulae older than 1 Gyr show [Fe/H]$>$-1.0$\pm$0.2
with very few exceptions (\citealp{idiart07}), whereas 
clusters analysed here show metallicities lower than [Fe/H]$<$-1.0. It
would be particularly interesting to carry out high resolution
spectroscopic analysis of individual stars in these clusters, in order
to check if there are very metal-poor clusters, apparently have
no counterpart among the planetary nebulae population, or at least very
few. The confirmation of metallicities of the most metal-poor
planetary nebulae would be needed.

Finally we identified a few clusters with ages between 1 and 8 to
10~Gyr (upper limits vary between code and SSP employed in our
calculations); therefore, we conclude that no clear age gap is present
in the SMC.


\begin{acknowledgements}  
PC is grateful to M. Koleva and P. Prugniel for the help with
\emph{ULySS} and to R. Cid-Fernandes for the long term help with {\sc
  Starlight}. BD, PC, BB, TI and LK acknowledge partial financial
support from the Brazilian agencies CNPq and Fapesp. PC acknowledges
the partial support of the EU through a Marie Curie Fellowship. The
authors are grateful to an anonymous referee for very helpful
suggestions.
The observations were carried out within Brazilian time in a ESO-ON
agreement and within an IAG-ON agreement funded by FAPESP project
n$^{\circ}$ 1998/10138-8. 
\end{acknowledgements}

\bibliographystyle{aa}
\bibliography{bibliography}


\appendix
\section{Spectral fits}
In this section, the best fits are shown for each cluster and Pegase-HR SSP models.

\begin{figure}[h]
\centering
\includegraphics[width=\columnwidth]{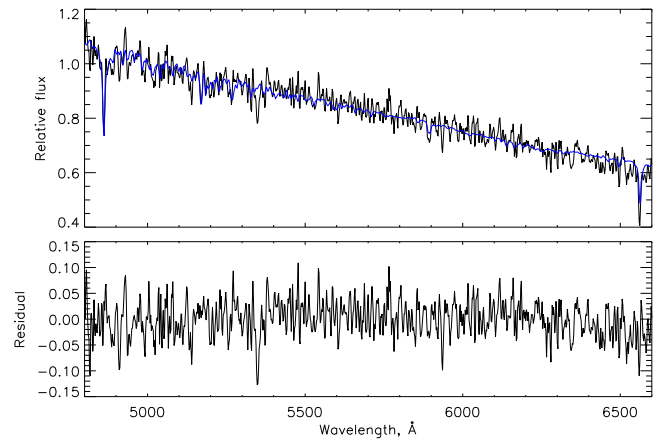}
\caption{\textit{Upper panel}: Observed spectrum (black line) for the cluster HW1 and the best model fitted by 
{\sc Starlight}+PEGASE-HR (blue line). \textit{Lower panel}: residuals of the fit.}
\label{fig_st_first}
\end{figure}

\begin{figure}[h]
\centering
\includegraphics[width=\columnwidth]{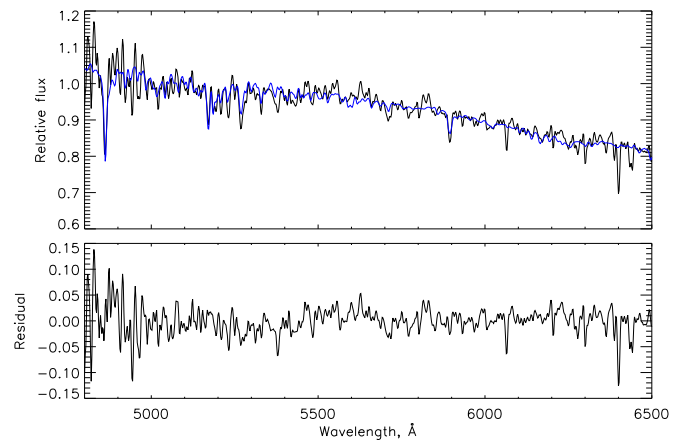}
\caption{Same as Fig. \ref{fig_st_first} for cluster K3 (ESO).}
\end{figure}

\begin{figure}[h]
\centering
\includegraphics[width=\columnwidth]{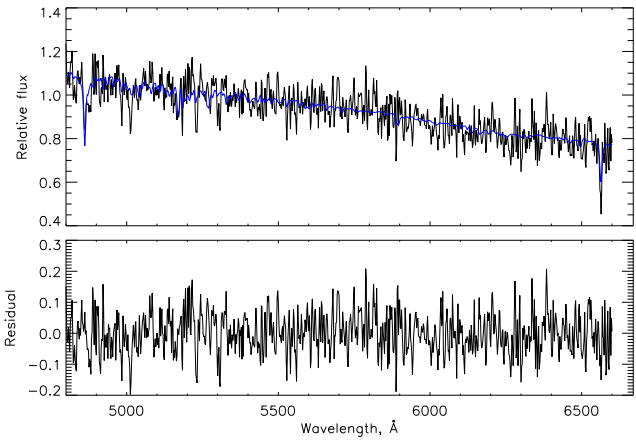}
\caption{Same as Fig. \ref{fig_st_first} for cluster K3 (LNA).}
\end{figure}

\begin{figure}[h]
\centering
\includegraphics[width=\columnwidth]{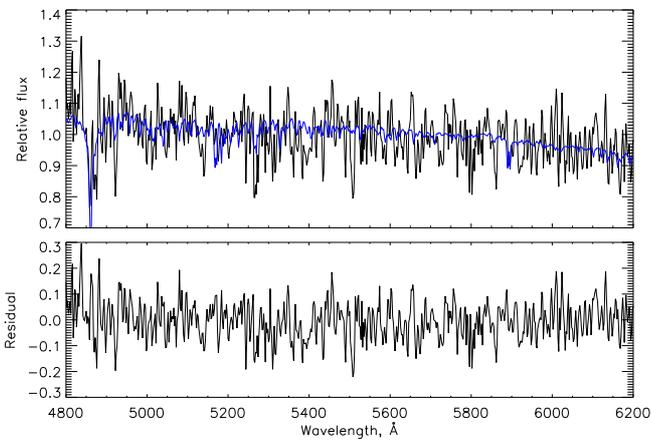}
\caption{Same as Fig. \ref{fig_st_first} for cluster L3.}
\end{figure}

\begin{figure}[h]
\centering
\includegraphics[width=\columnwidth]{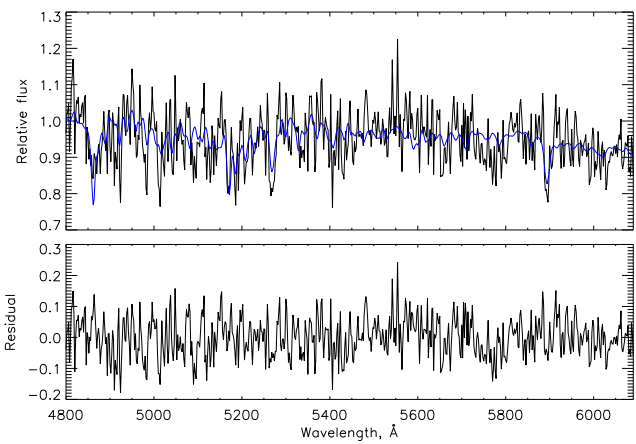}
\caption{Same as Fig. \ref{fig_st_first} for cluster L11.}
\end{figure}

\begin{figure}[h]
\centering
\includegraphics[width=\columnwidth]{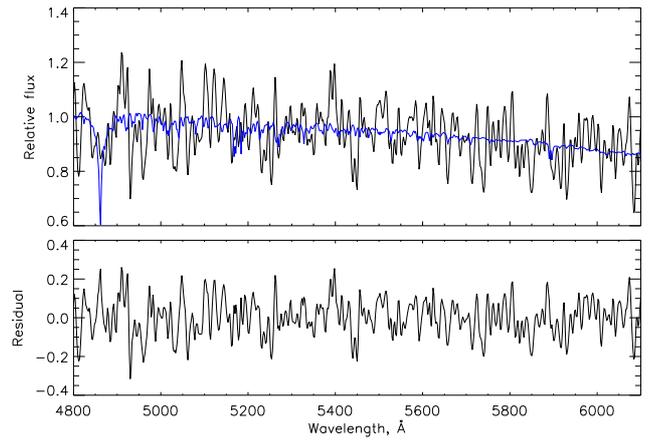}
\caption{Same as Fig. \ref{fig_st_first} for cluster L113.}
\end{figure}

\begin{figure}[h]
\centering
\includegraphics[width=\columnwidth]{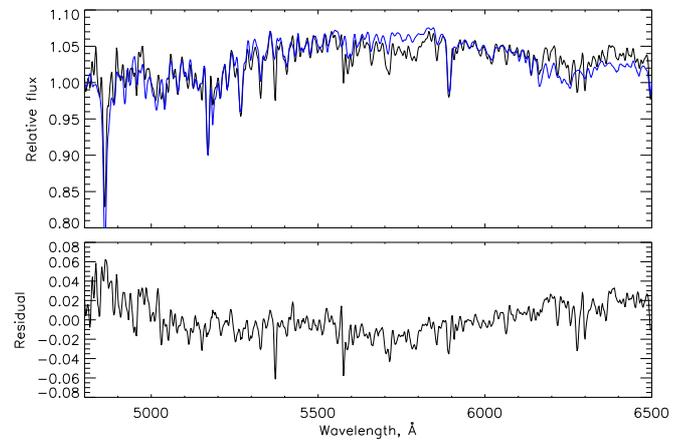}
\caption{Same as Fig. \ref{fig_st_first} for cluster NGC~121 (ESO).}
\end{figure}

\begin{figure}[h]
\centering
\includegraphics[width=\columnwidth]{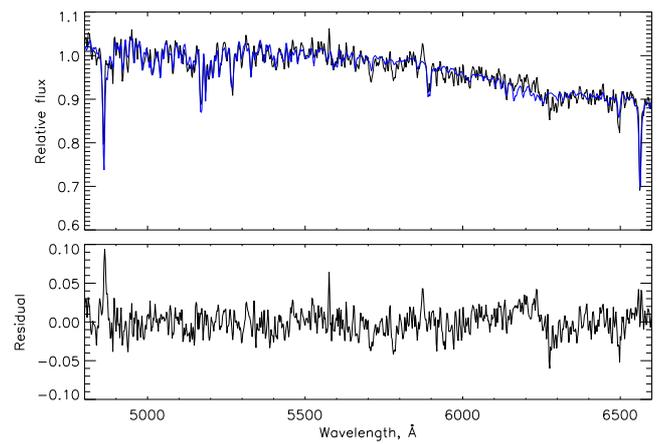}
\caption{Same as Fig. \ref{fig_st_first} for NGC~121 (LNA).}
\end{figure}

\begin{figure}[h]
\centering
\includegraphics[width=\columnwidth]{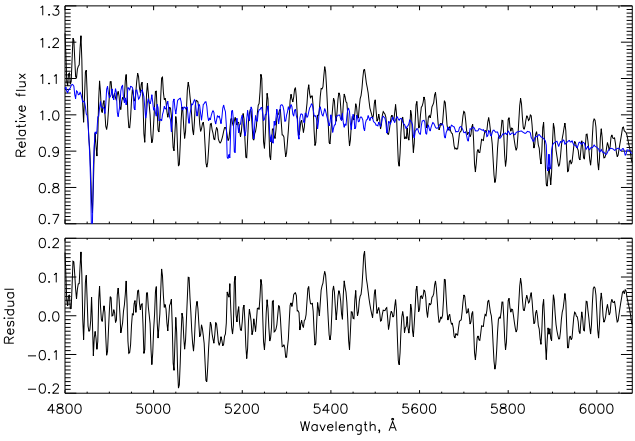}
\caption{Same as Fig. \ref{fig_st_first} for cluster NGC~152.}
\end{figure}

\begin{figure}[h]
\centering
\includegraphics[width=\columnwidth]{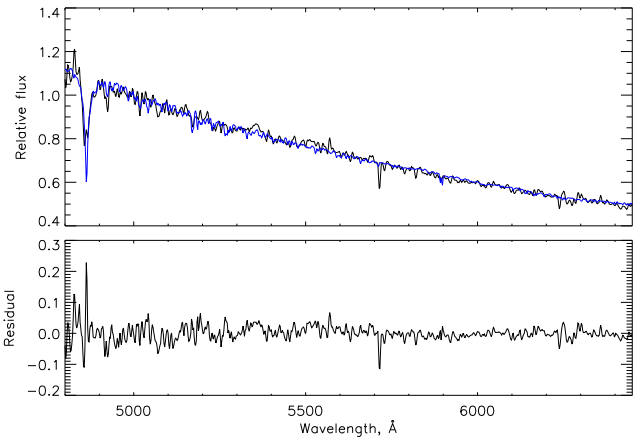}
\caption{Same as Fig. \ref{fig_st_first} for cluster NGC~222.}
\end{figure}

\begin{figure}[h]
\centering
\includegraphics[width=\columnwidth]{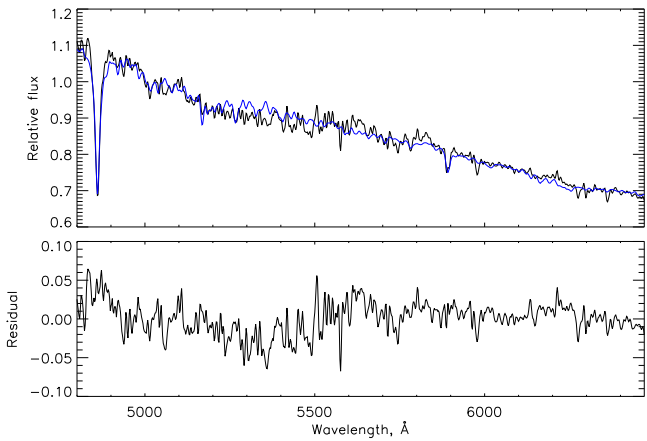}
\caption{Same as Fig. \ref{fig_st_first} for cluster NGC~256.}
\end{figure}

\begin{figure}[h]
\centering
\includegraphics[width=\columnwidth]{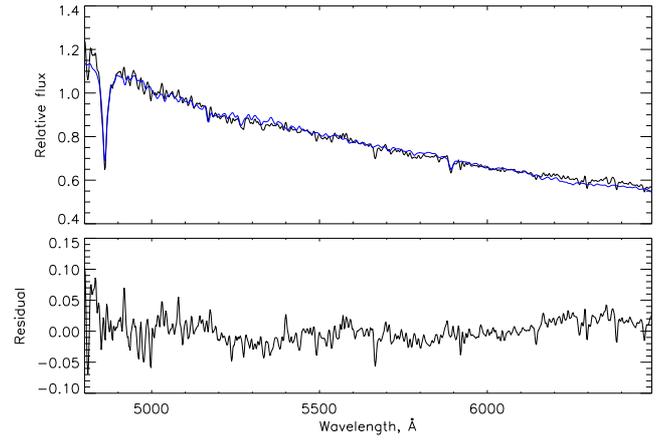}
\caption{Same as Fig. \ref{fig_st_first} for cluster NGC~269.}
\end{figure}

\begin{figure}[h]
\centering
\includegraphics[width=\columnwidth]{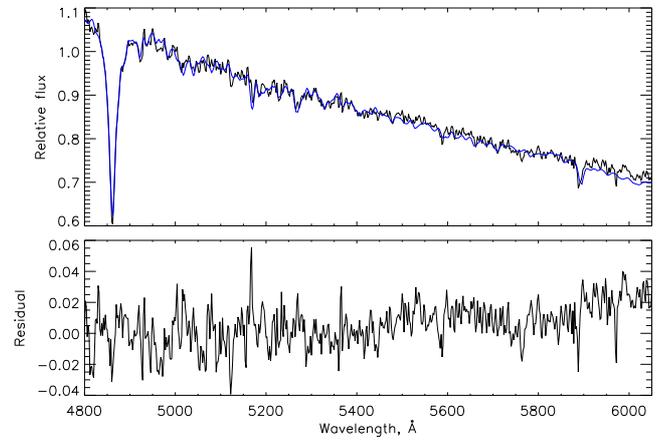}
\caption{Same as Fig. \ref{fig_st_first} for cluster NGC~294.}
\end{figure}

\begin{figure}[h]
\centering
\includegraphics[width=\columnwidth]{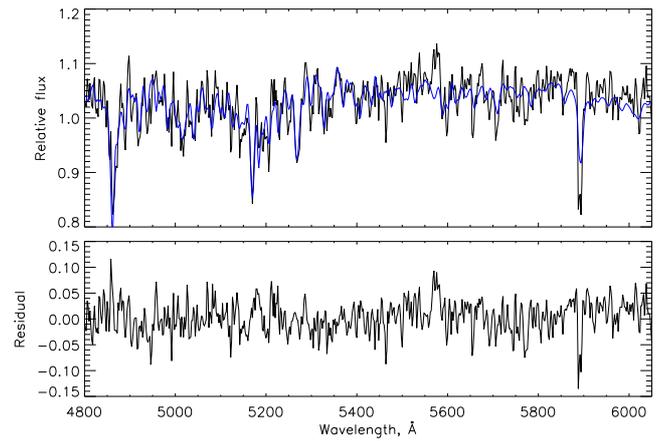}
\caption{Same as Fig. \ref{fig_st_first} for cluster NGC~361 (ESO99).}
\end{figure}

\begin{figure}[h]
\centering
\includegraphics[width=\columnwidth]{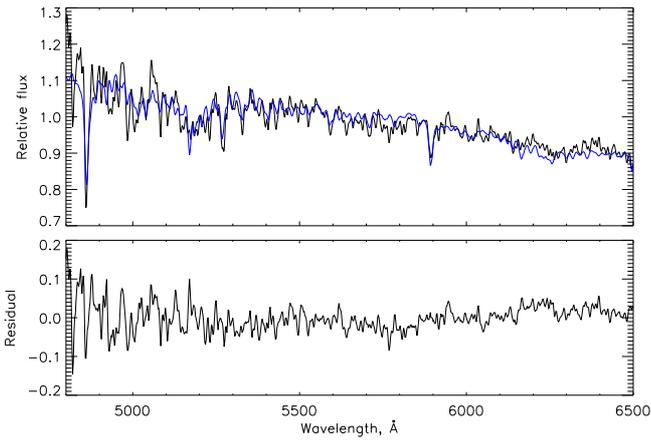}
\caption{Same as Fig. \ref{fig_st_first} for NGC~361 (ESO00).}
\end{figure}

\begin{figure}[h]
\centering
\includegraphics[width=\columnwidth]{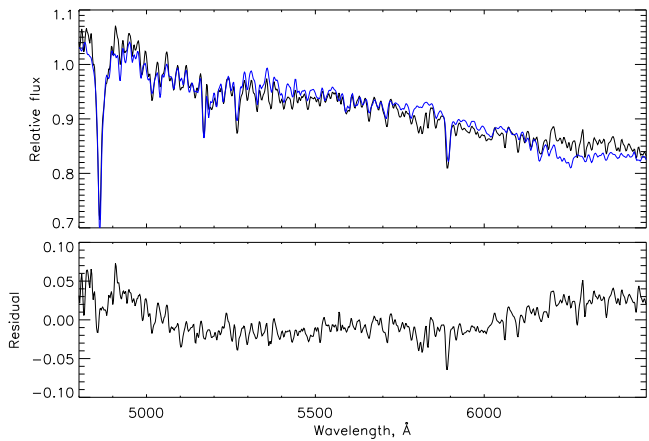}
\caption{Same as Fig. \ref{fig_st_first} for cluster NGC~419.}
\end{figure}

\begin{figure}[h]
\centering
\includegraphics[width=\columnwidth]{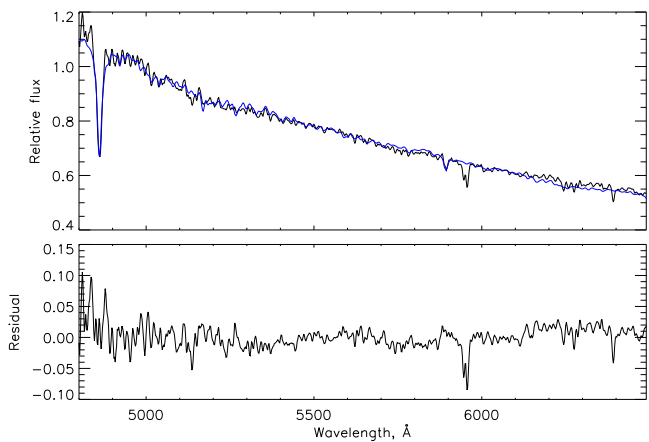}
\caption{Same as Fig. \ref{fig_st_first} for cluster NGC~458.}
\label{fig_st_last}
\end{figure}

\begin{figure}[h]
\centering
\includegraphics[width=\columnwidth]{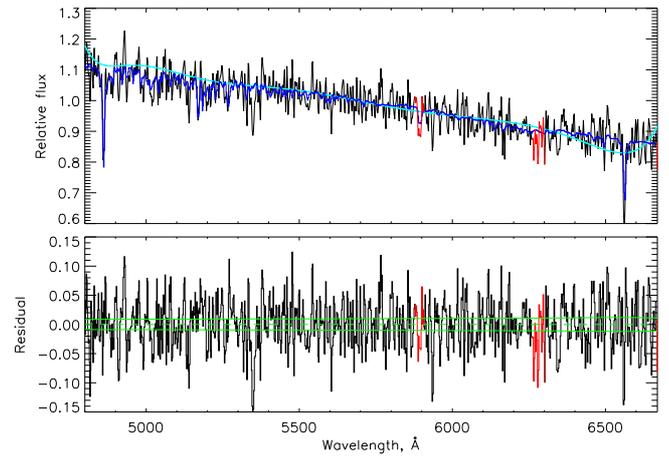}
\caption{\textit{Upper panel}: Observed spectrum (black line) for the cluster HW1 and the best model fitted by 
\emph{ULySS}+PEGASE-HR (blue line). Red pixels correspond to regions of telluric lines and were rejected from the fit. \textit{Lower panel}: residuals of the fit. The continuous green lines mark the 1-$\sigma$ deviation.}
\label{fig_uly_first}
\end{figure}

\clearpage

\begin{figure}[h]
\centering
\includegraphics[width=\columnwidth]{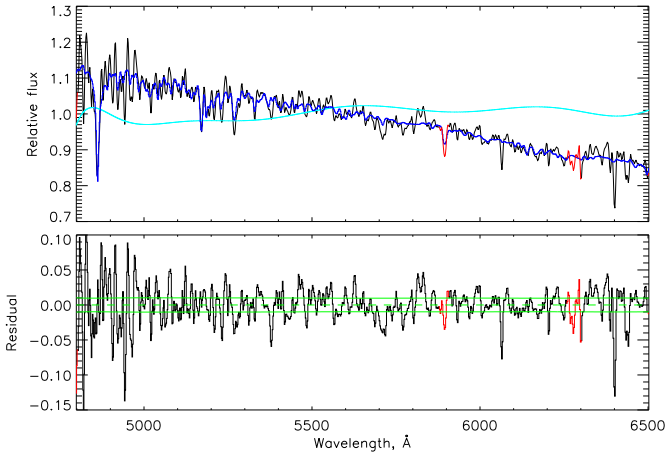}
\caption{Same as Fig. \ref{fig_uly_first} for cluster K3 (ESO).}
\end{figure}

\begin{figure}[h]
\centering
\includegraphics[width=\columnwidth]{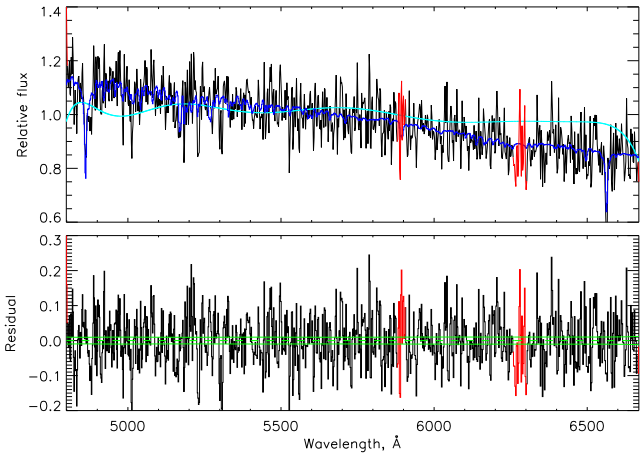}
\caption{Same as Fig. \ref{fig_uly_first} for cluster K3 (LNA).}
\end{figure}

\begin{figure}[h]
\centering
\includegraphics[width=\columnwidth]{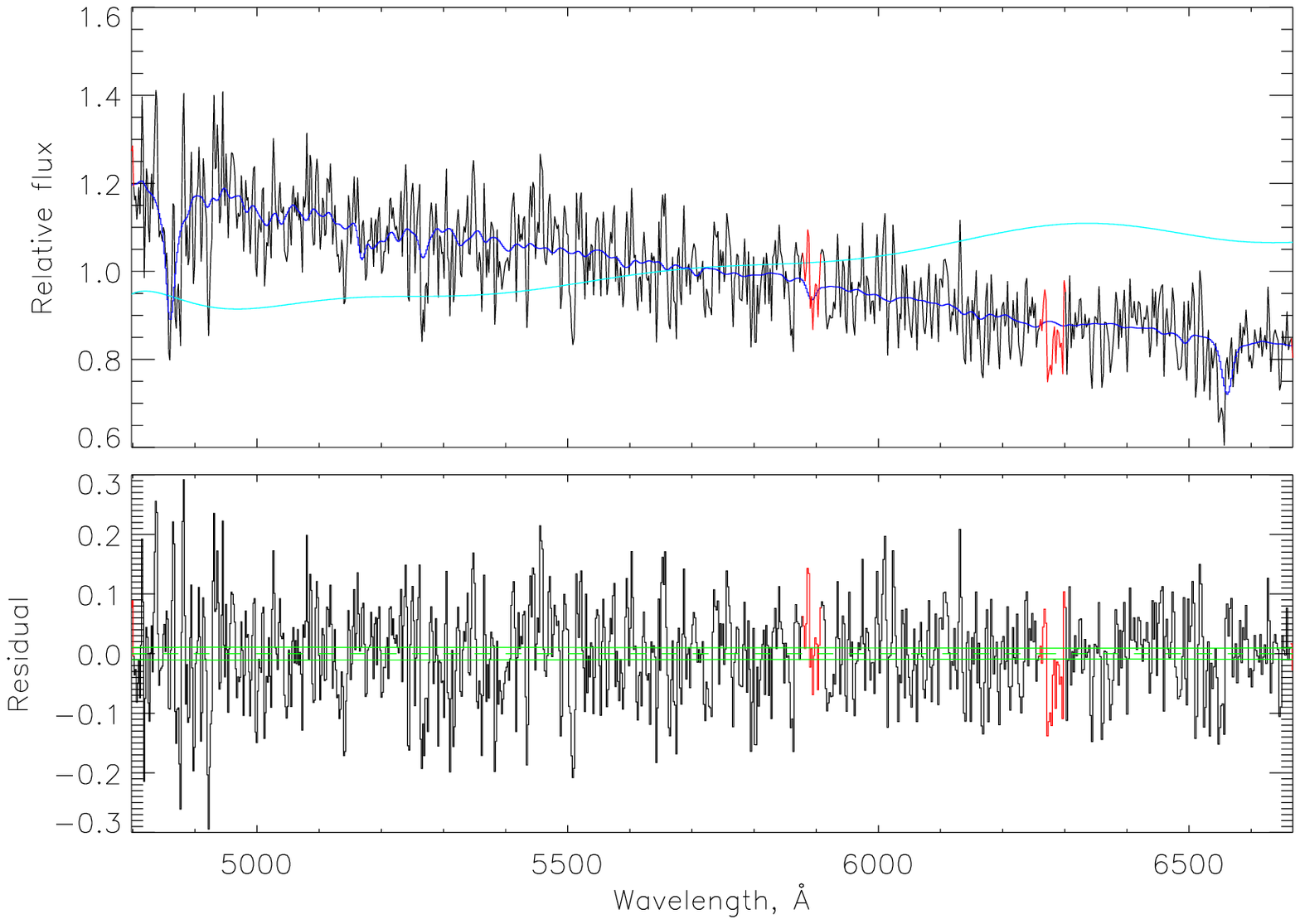}
\caption{Same as Fig. \ref{fig_uly_first} for cluster L3.}
\end{figure}

\begin{figure}[h]
\centering
\includegraphics[width=\columnwidth]{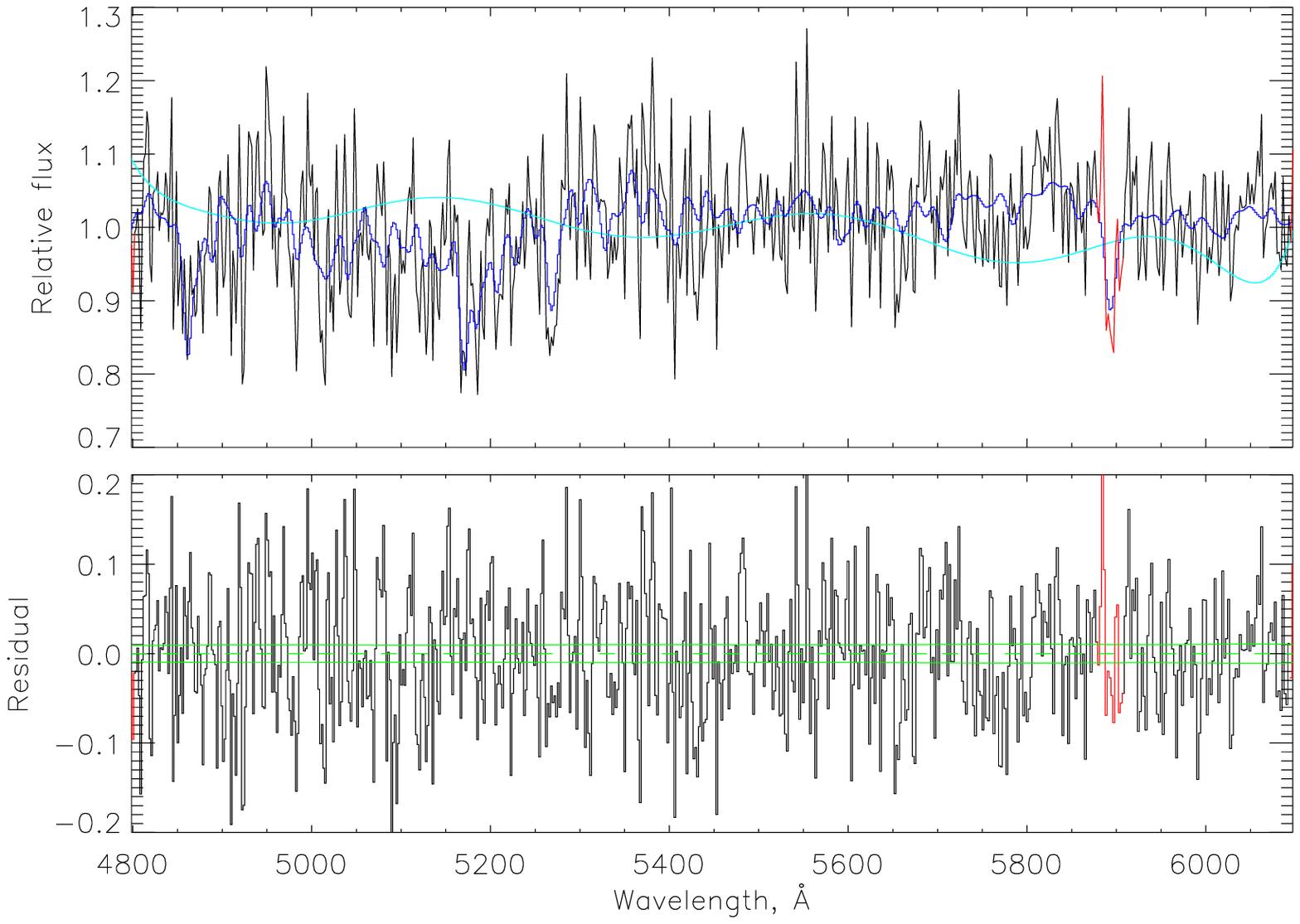}
\caption{Same as Fig. \ref{fig_uly_first} for cluster L11.}
\end{figure}

\begin{figure}[h]
\centering
\includegraphics[width=\columnwidth]{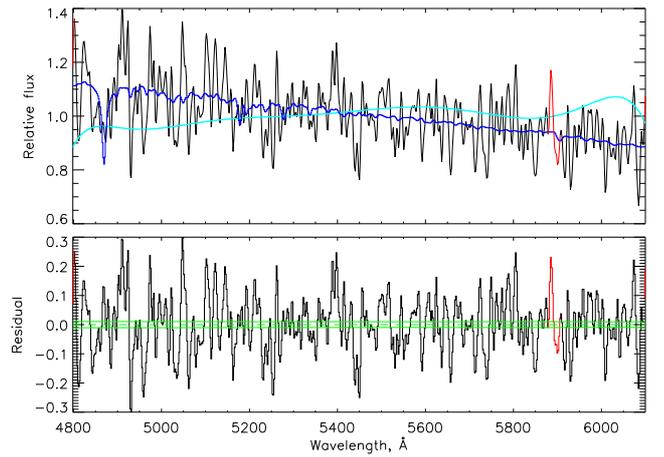}
\caption{Same as Fig. \ref{fig_uly_first} for cluster L113.}
\end{figure}

\begin{figure}[h]
\centering
\includegraphics[width=\columnwidth]{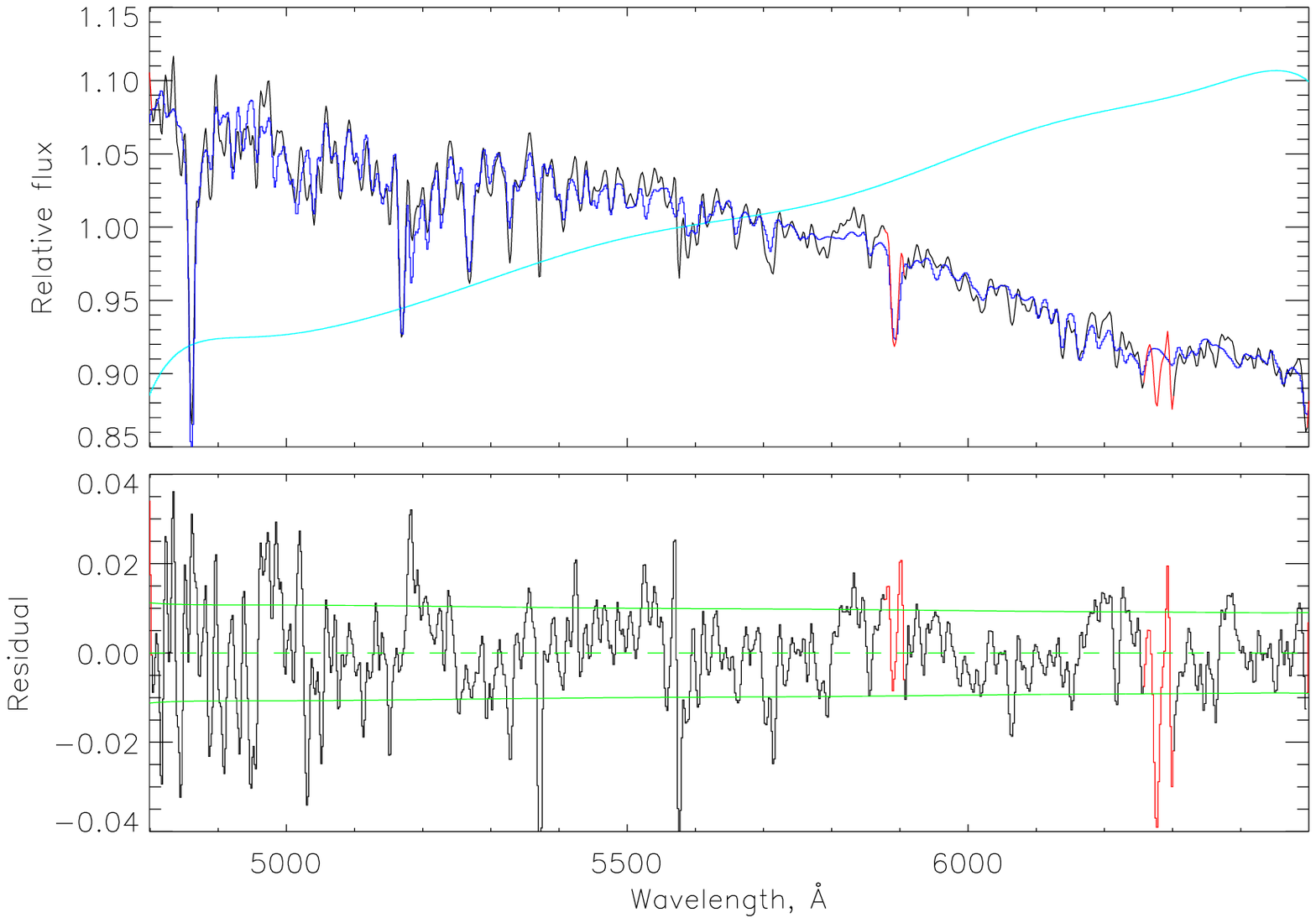}
\caption{Same as Fig. \ref{fig_uly_first} for cluster NGC~121 (ESO).}
\end{figure}

\begin{figure}[h]
\centering
\includegraphics[width=\columnwidth]{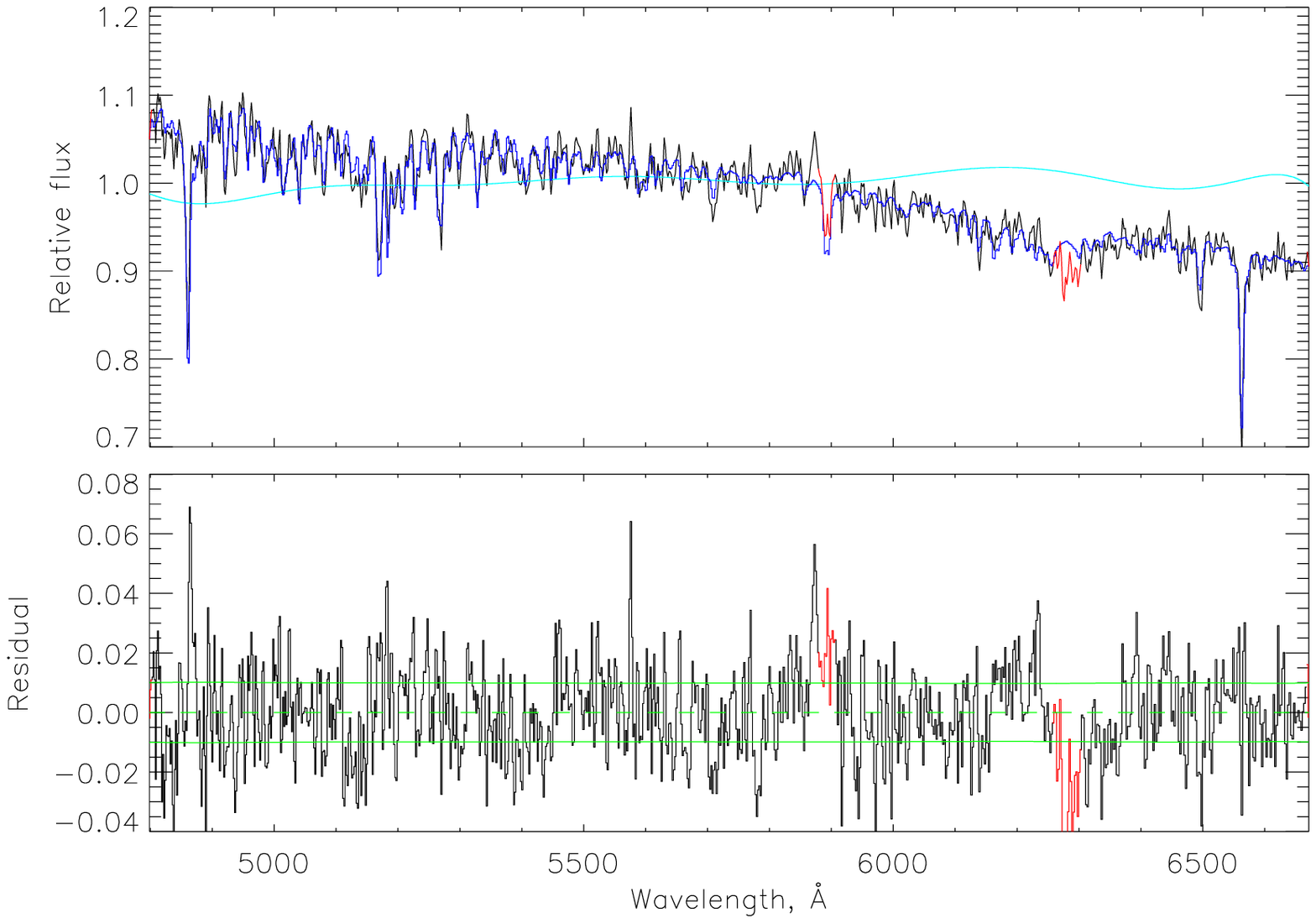}
\caption{Same as Fig. \ref{fig_uly_first} for NGC~121 (LNA).}
\end{figure}

\begin{figure}[h]
\centering
\includegraphics[width=\columnwidth]{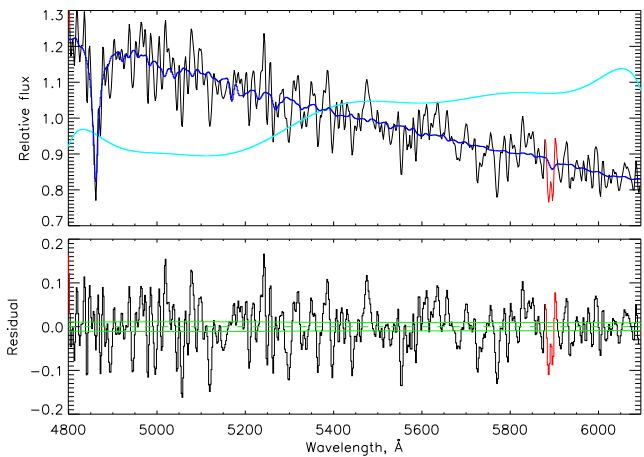}
\caption{Same as Fig. \ref{fig_uly_first} for cluster NGC~152.}
\end{figure}

\begin{figure}[h]
\centering
\includegraphics[width=\columnwidth]{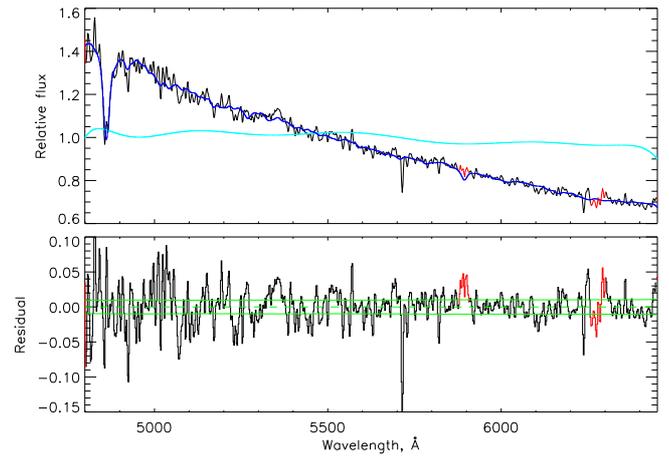}
\caption{Same as Fig. \ref{fig_uly_first} for cluster NGC~222.}
\end{figure}

\begin{figure}[h]
\centering
\includegraphics[width=\columnwidth]{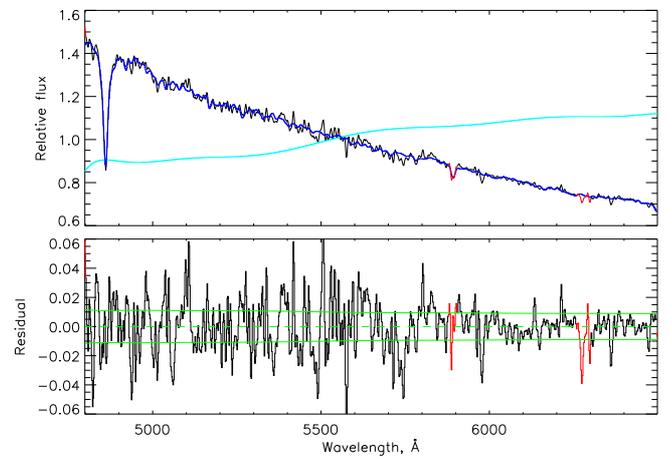}
\caption{Same as Fig. \ref{fig_uly_first} for cluster NGC~256.}
\end{figure}

\begin{figure}[h]
\centering
\includegraphics[width=\columnwidth]{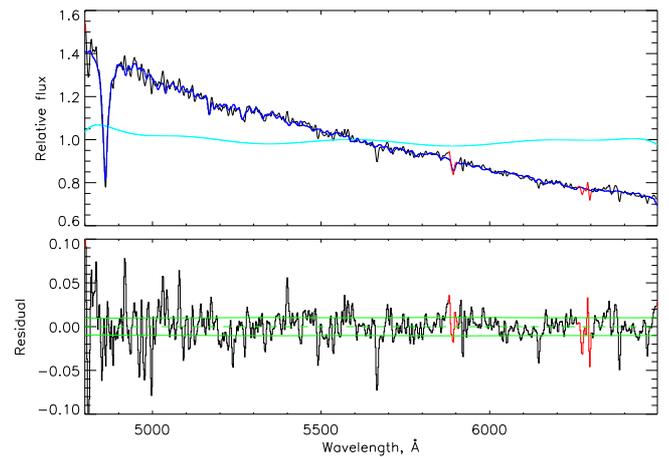}
\caption{Same as Fig. \ref{fig_uly_first} for cluster NGC~269.}
\end{figure}

\begin{figure}[h]
\centering
\includegraphics[width=\columnwidth]{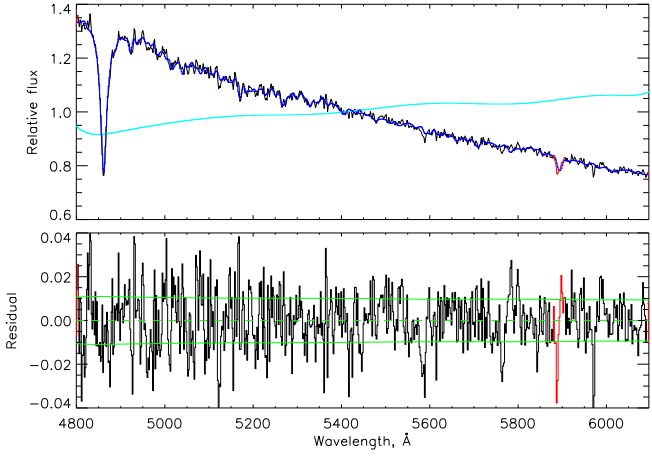}
\caption{Same as Fig. \ref{fig_uly_first} for cluster NGC~294.}
\end{figure}

\begin{figure}[h]
\centering
\includegraphics[width=\columnwidth]{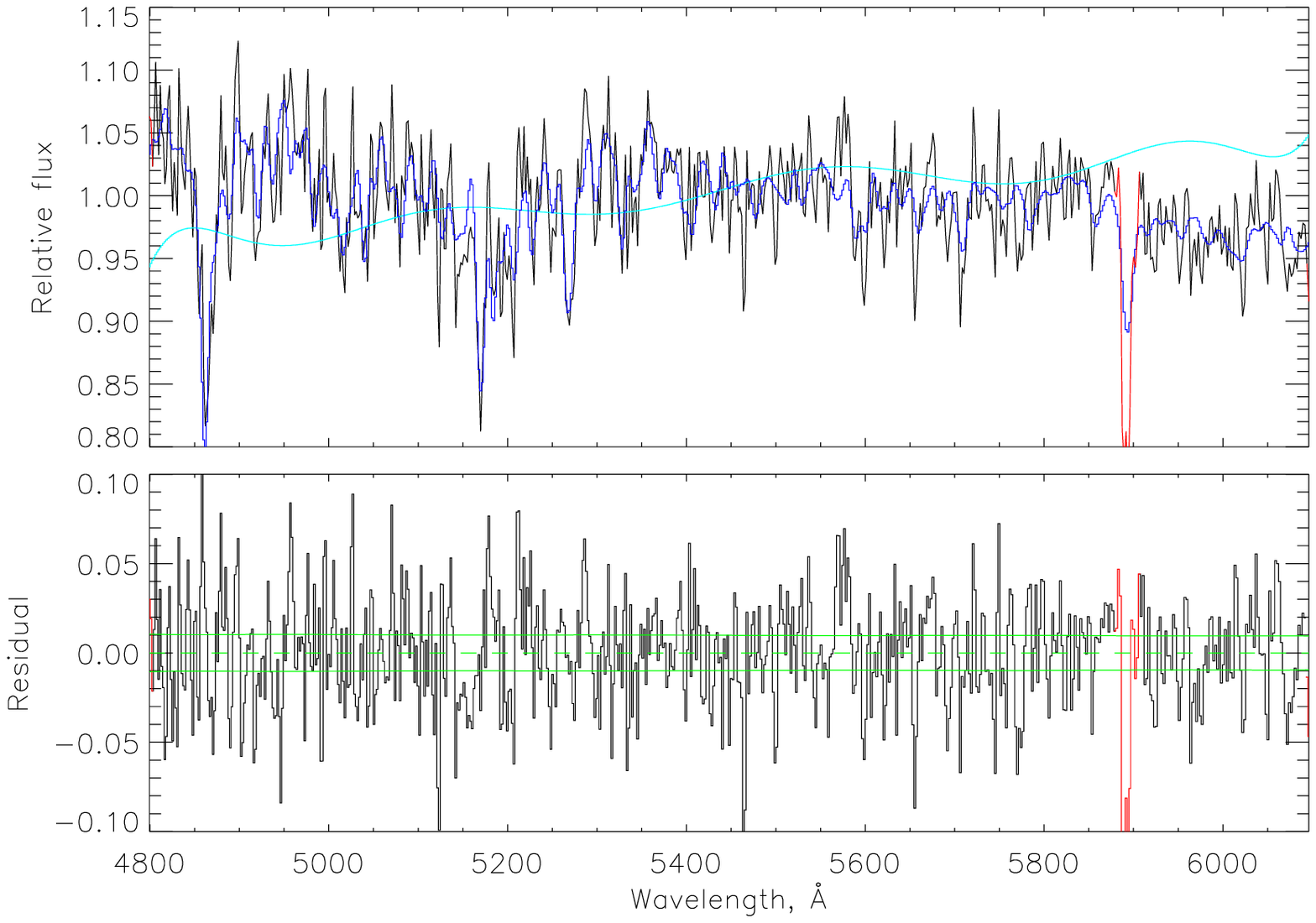}
\caption{Same as Fig. \ref{fig_uly_first} for cluster NGC~361 (ESO99).}
\end{figure}

\begin{figure}[h]
\centering
\includegraphics[width=\columnwidth]{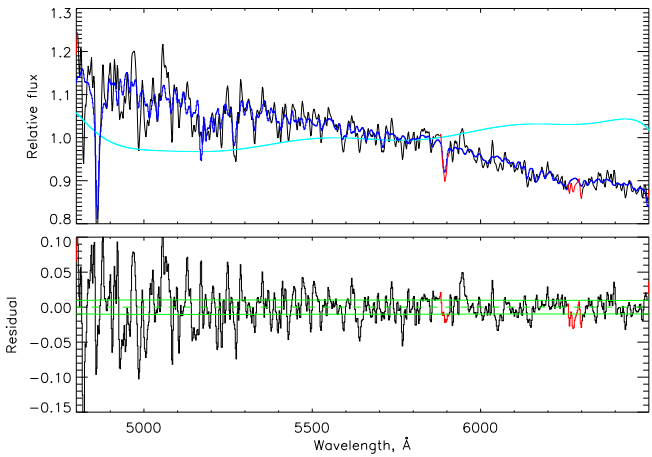}
\caption{Same as Fig. \ref{fig_uly_first} for NGC~361 (ESO00).}
\end{figure}

\begin{figure}[h]
\centering
\includegraphics[width=\columnwidth]{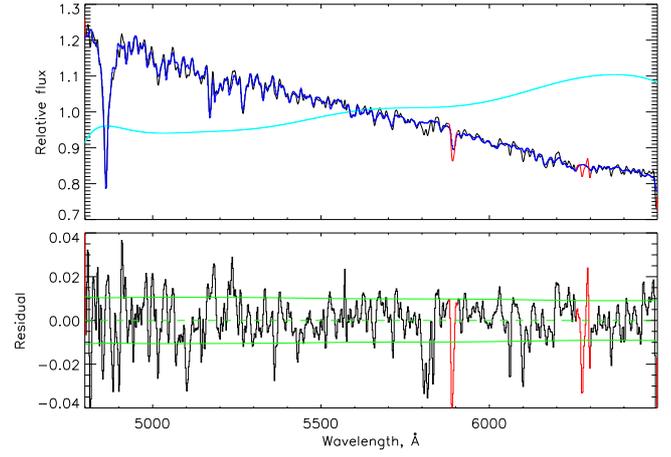}
\caption{Same as Fig. \ref{fig_uly_first} for cluster NGC~419.}
\end{figure}

\begin{figure}[h]
\centering
\includegraphics[width=\columnwidth]{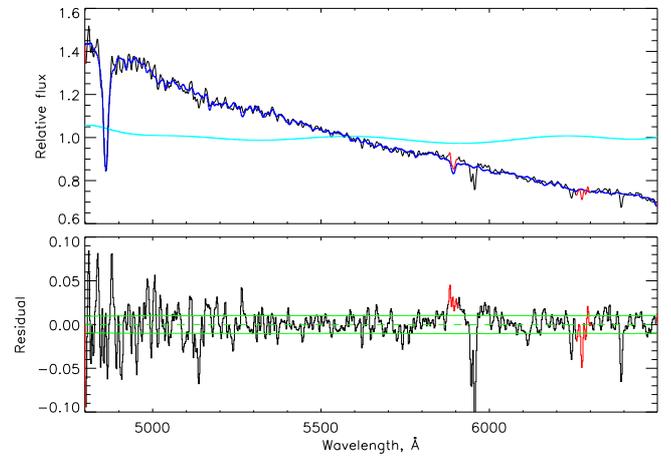}
\caption{Same as Fig. \ref{fig_uly_first} for cluster NGC~458.}
\label{fig_uly_last}
\end{figure}

\end{document}